\documentclass[12pt,a4,sans]{article}
\usepackage[latin1]{inputenc}
\usepackage{amssymb}
\usepackage{amsmath}
\usepackage{amsthm}
\usepackage{amsfonts}
\usepackage[pdftex]{graphicx}
\usepackage{geometry}
\usepackage{braket}
\usepackage{enumerate}
\usepackage[usenames,dvipsnames]{xcolor}
\usepackage{setspace}
\usepackage[backgroundcolor=white,linecolor=black]{todonotes}
\usepackage[in]{fullpage}
\usepackage{hyperref}
\usepackage{array}
\usepackage{cancel}
\usepackage{wrapfig}
\usepackage[font=small,labelfont=bf]{caption}
\usepackage{anysize}

\numberwithin{equation}{section}

\marginsize{2.4cm}{2.4cm}{2cm}{2cm}

\hypersetup{
colorlinks=true,
linktoc=page,
citecolor=Blue,
linkcolor=Blue,
urlcolor=Blue} 

\makeatletter % Need for anything that contains an @ command 
\renewcommand{\maketitle} % Redefine maketitle to conserve space
 { \begingroup \begin{center} \large {\bf \@title}
 	\vskip 5pt \large \@author \\ \vskip 5pt \@date \end{center}
   \vskip 5pt \endgroup \setcounter{footnote}{0} }
\makeatother % End of region containing @ commands

\newcommand{\pic}[2]{\vcenter{\hbox{\includegraphics[scale=#1]{#2}}}}

\newcommand{\comments}[1]{}
\newcommand{\la}{\langle}
\newcommand{\ra}{\rangle}

\newcommand{\Tr}{\text{Tr}}
\newcommand{\MHVb}{\overline{\text{MHV}}}
\renewcommand{\b}[1]{\braket{#1}}

\renewcommand{\O}{\mathcal{O}}

\newcommand{\be}{\begin{equation}}
\newcommand{\ee}{\end{equation}}

\newcommand{\nn}{\nonumber\\}

\def\beqa{\begin{eqnarray}}
\def\eeqa{\end{eqnarray}}
\def\beq{\begin{equation}}
\def\eeq{\end{equation}}

\def\Tr{{\rm Tr}}
\def\one{\mbox{1 \kern-.59em {\rm l}}}

\def\cD{{\cal D}}  
  
 \def\cK{{\cal K}} \def\cL{{\cal L}}
 \def\cN{{\cal N}} \def\cO{{\cal O}}
  \def\cR{{\cal R}}

 \def\cZ{{\cal Z}}

\def\uno{\mbox{1 \kern-.59em {\rm l}}}

\def\one{1\!\!1\,\,}

\def\bcomment#1{}
\newcommand{\ls}{\left[}
\newcommand{\rs}{\right]}

\def\eps{\epsilon}

\usepackage{slashed}
\usepackage{caption}

\usepackage[nottoc,notlot,notlof]{tocbibind}
\usepackage[nosort]{cite}
\usepackage{pstricks}         
\usepackage{color}
\usepackage{parskip}
\graphicspath{{./Images/}}

\long\def\symbolfootnote[#1]#2{\begingroup%
\def\thefootnote{\fnsymbol{footnote}}\footnote[#1]{#2}\endgroup}

\begin{document}

\begin{flushright}
QMUL-PH-16-11\\
HU-EP-16/21\\
HU-MATH-2016-13
\end{flushright}

\vspace{20pt}

\begin{center}

{\Large \bf The  $SU(2|3)$ dynamic two-loop form factors}\\
%\vspace{0.3 cm}
%{\Large \bf in form factors  }

%
\vspace{45pt}
%\end{center}

{\mbox {\bf 
\hspace{-1cm} 
A.~Brandhuber$^{\,\sharp}$, M.~Kostaci\'{n}ska$^{\,\sharp}$, B.~Penante$^{\,\sharp,\, \flat}$, G.~Travaglini$^{\,\sharp}$ and D.~Young$^{\,\sharp}$}}%
\symbolfootnote[4]{
{\tt  \{ \tt \!\!\!\!\!\! a.brandhuber,  m.m.kostacinska, b.penante, g.travaglini, d.young\}@qmul.ac.uk}
}

\vspace{0.5cm}
\begin{center}
{\small \em
$^{\sharp}$ Centre for Research in String Theory\\
School of Physics and Astronomy\\
Queen Mary University of London\\
Mile End Road, London E1 4NS, UK
}
\end{center}
\vspace{0.5cm}
\begin{center}
{\small \em
$^{\flat}$ Institut f\"{u}r Physik und IRIS Adlershof\\
Humboldt Universit\"{a}t zu Berlin\\
Zum Gro\ss en Windkanal 6, 12489 Berlin, Germany
}
\end{center}
%\vspace{-8pt}

\vspace{40pt}

{\bf Abstract}
\end{center}

\vspace{0.3cm} 

\noindent

\noindent
We compute two-loop form factors of operators in the $SU(2|3)$ closed subsector of \mbox{$\cN\!=\!4$} supersymmetric Yang-Mills. In particular, we focus on the non-protected,  dimension-three  operators $\Tr (X[Y,Z])$ and $\Tr ( \psi \psi)$ for which we  compute the four possible two-loop form factors, and corresponding remainder functions,  with external states $\langle \bar{X} \bar{Y} \bar{Z}|$ and $\langle \bar{\psi} \bar{\psi}|$. 
Interestingly, the maximally transcendental part of the two-loop  remainder of $\langle\bar{X} \bar{Y} \bar{Z}| \Tr (X[Y,Z]) |0\rangle$ turns out to be identical to that of the corresponding known quantity for the half-BPS operator $\Tr (X^3)$. We also find a surprising connection between the terms subleading in transcendentality and certain a priori unrelated remainder densities introduced in the study of the  spin chain Hamiltonian in the $SU(2)$ sector. Next, we use our calculation to  resolve the mixing, recovering anomalous dimensions and eigenstates of the dilatation operator in the $SU(2|3)$ sector at two loops. 
We also speculate on potential connections between our calculations in $\cN\!=\!4$ super Yang-Mills and Higgs + multi-gluon amplitudes in QCD  in an effective Lagrangian approach.

\setcounter{page}{0}
\thispagestyle{empty}
\newpage

%%%%%%%%%%%%%%%%%% TABLE OF CONTENTS %%%%%%%%%%%%%%%%%%%%%%%%%%%%%%%%%

\setcounter{tocdepth}{4}
\hrule height 0.75pt
\tableofcontents
\vspace{0.8cm}
\hrule height 0.75pt
\vspace{1cm}

\setcounter{tocdepth}{2}

\newpage
%%%%%%%%%%%%%%%%%%%%%%%%%%%%%%%%%%%%%%%%%%%%%%%%%%%%%%%%%%%

\section{Introduction}

The study of form factors of composite operators is a very active area of research. After the pioneering paper \cite{vanNeerven:1985ja}, interest in the calculation of form factors in supersymmetric theories was rekindled at strong coupling in  
\cite{Alday:2007he} and  at weak coupling in \cite{Brandhuber:2010ad}. Specifically, in   \cite{Brandhuber:2010ad} the study of the simplest possible  form factors was undertaken, namely form factors  of quadratic half-BPS operators in $\cN\!=\!4$ supersymmetric Yang-Mills (SYM). 
The three- and four-loop result for the simplest, two-point (or Sudakov) form factors were then derived in 
\cite{Gehrmann:2011xn, Boels:2015yna}, respectively. 
In \cite{Brandhuber:2012vm}    two-loop  form factors of the form  $\langle \bar{X}\bar{X} g^{\pm} | \Tr (X^2) | 0\rangle$
were computed, 
where $X$ is one of the three complex scalar fields of the theory, and $g^+$ ($g^-$) denotes a gluon of positive (negative) helicity. In that paper  it was also shown that  these form factors are identical  to   the  form factors of the self-dual field strength%
\footnote{Or, more precisely, of the on-shell Lagrangian.} 
$F_{\rm SD}$,  $\langle g^+ g^+ g^\pm | \Tr (F_{\rm SD})^2 | 0\rangle$ (divided by the corresponding tree-level contribution)   thanks to supersymmetric Ward identities 
\cite{Brandhuber:2011tv, Bork:2011cj}.  Indeed, the operators $\Tr (X^2)$ and the on-shell Lagrangian belong to the simplest operator multiplet in the  $\cN\!=\!4$ theory, namely the protected stress-tensor multiplet. 
Remarkably, in QCD the   form factors  of $\Tr (F^2)$ compute the leading contribution to  Higgs + multi-gluon amplitudes in an effective Lagrangian approach \cite{Wilczek:1977zn, Shifman:1979eb, Dawson:1990zj, Spira:1995rr,Dixon:2004za,Badger:2004ty,Badger:2006us,Badger:2007si,Badger:2009hw} in the large   top mass limit. The corresponding interaction   has the form 
$\cL^{(0)}_{\rm eff}\!\sim\!H \, \Tr (F^2)$, and hence the quantity%
\footnote{Recall that  we can separate out $F^2 = F_{\rm SD}^2 + F_{\rm ASD}^2$.}
\beq
\label{inizio} 
\langle gg\cdots g  | \int\!d^4x \ e^{ - i q\cdot x} \  \Tr (F^2) (x) | 0 \rangle
\ 
\eeq
precisely computes the amplitude for the process $H\to gg \cdots g$, with $q^2 = m_{\rm H}^2$. 

Of course, there is no a priori connection between the quantity  \eqref{inizio} evaluated  in QCD and in $\cN\!=\!4$ SYM. 
Yet, in  \cite{Brandhuber:2012vm} it was realised  that the three-point form factor computed there, $\langle \bar{X}\bar{X} g^+ | \Tr (X^2) | 0\rangle$,  is identical  to the maximally transcendental part of the amplitudes for $H\to g^+ g^+ g^\pm$ calculated in \cite{Gehrmann:2011aa, Koukoutsakis:2003nba}. This led to the conjecture that the ``most complicated part", i.e.~the maximally transcendental contribution to  Higgs plus multi-gluon processes, at infinite top quark mass,  can  in fact be computed  using $\cN\!=\!4$ SYM. 

The coupling $\cL^{(0)}_{\rm eff}$ quoted earlier is only the first in an effective Lagrangian description of gluon fusion processes. Subleading corrections have been studied in  a number of papers, see e.g.~\cite{Neill:2009tn,Dawson:2014ora}, where the expansion of the effective Lagrangian  is written as 
\beq
\label{corre}
\cL_{\rm eff} \ = \ \hat{C}_0 O_0 \, + \, {1\over m_{\rm top}^2} \sum_{i=1}^4 \hat{C}_i O_i \, + \, \cO
\left({1\over m_{\rm top}^4} \right)\ , 
\eeq
where $O_i$, $i=1, \ldots , 4$ are dimension-7 operators and $O_0 = H \, \Tr (F^2)$. 

Some of the operators in the set $\{O_i\}_{i=1}^4$ do not contain quarks, and as such can be considered also in $\cN\!=\!4$ SYM. 
In this paper we would like to suggest the relevance of computing form factors of such operators in the maximally supersymmetric theory, and comparing to the QCD results. One possible very interesting scenario is that the $\cN\!=\!4$ SYM calculation continues to capture the maximally transcendental part of the corresponding QCD calculation. 
In particular, the following two operators can be considered, 
\beq
O_1 :=  H\, \Tr (F^3) =  H\, \big[ \Tr (F_{\rm SD} ^3)  + \Tr (F_{\rm ASD} ^3)\big]\ , \qquad   O_2 := H\, \Tr \big[ (D^\mu F^{\nu \rho})(D_\mu F_{\nu \rho})\big]
\ , 
\eeq
which in  QCD are both multiplicatively renormalisable at one loop \cite{Gracey:2002he}. 
Let us briefly discuss the case of $O_1$, and in particular  the form factor $\langle g^+ g^+ g^+ | \Tr (F^3) | 0 \rangle$. At tree-level and zero momentum transfer (i.e. $q=0$, where $q$ is the momentum carried by the operator), 
 these form factors become amplitudes produced by higher-dimensional couplings, and  have been considered recently in 
 \cite{Dixon:2004za, Broedel:2012rc}. At $q\neq 0$, they have been studied at tree-level and one loop in \cite{Dawson:2014ora}. In $\cN\!=\!4$ SYM and at one loop,  it turns out that the operator $\Tr (F^3)$ has the same anomalous dimension as the Konishi operator 
  (the calculation of the three-gluon form factor for this operator is currently under investigation). A technically simpler, but equally  interesting computation consists of focusing  on simpler operators, still containing three fields, and several candidate operators immediately come to mind. 
 The half-BPS operator 
 \beq
 \cO_{\rm BPS} \ = \ 
 \Tr (X^3)
 \ , 
 \eeq 
 and its form factors have been studied at one and two loops in 
 \cite{Penante:2014sza, Brandhuber:2014ica}. A priori it is however too simple -- for instance, unlike $\Tr (F^3)$ in QCD, $ \cO_{\rm BPS}$ is protected. Scalar fields are of course preferred, as their form factors are the simplest possible. In order to get a non-protected, trilinear operator we need to consider three complex scalar fields, which we can choose to~be 
 \begin{align}
\label{eq:XYZ-phi}
\begin{split}
{X}\,:=\,\phi_{12}\,,\quad {Y}\,:=\,\phi_{23}\,,\quad {Z}\,:=\,\phi_{31} \ . 
\end{split}
\end{align}
From these fields, one can immediately construct the operators
\beqa
\label{obps}
\tilde{\cO}_{\rm BPS} &:= &\Tr(X\{Y,Z\})\ ,
%\cO_{\rm BPS} &:= &\Tr(X\{Y,Z\})\ , 
\\
\label{eq:OB}
 \cO_{B} &:=& \Tr(X[Y,Z])  \ . 
\eeqa 
While the first operator is another half-BPS combination,%
\footnote{It is symmetric and traceless once written in $SO(6)$ indices.}
quantum corrections lead to mixing between $\cO_B$ and the dimension-three operator, 
\begin{align}
\begin{split}
\cO_F := {1\over 2} \Tr (\psi^\alpha \psi_\alpha)\ , 
\end{split}
\end{align}
 where we have defined 
\beq
\label{eq:psi}
\psi_{\alpha}:=\psi_{123, \alpha} \ . 
\eeq
The fields $\{ \phi_{12}, \phi_{23}, \phi_{31}; \psi_{123, \alpha}\}$ are precisely the letters of the 
$SU(2|3)$ closed subsector of $\cN\!=\!4$ SYM.  It has been studied by Beisert in \cite{Beisert:2003jj, Beisert:2003ys}, where the dilatation operator was determined up to three loops. Apart from being closed under operator mixing, there is another important feature  of this sector: it gives rise to length-changing interactions in the dilatation operator, such as $XYZ \leftrightarrow \psi \psi$, unlike the (simpler) $SU(2)$ sector.

Motivated by the above discussion, we now describe in more detail  the goals of this paper. In the following  we will focus on the non-protected, (classically) dimension-three  operators $\O_B$ and $\O_F$ for which we  compute the four possible two-loop form factors, and corresponding remainder functions  with external states $\langle \bar{X} \bar{Y} \bar{Z}|$ and $\langle \bar{\psi} \bar{\psi}|$. It is convenient, and natural from the point of view of operator mixing discussed later, to package them into a matrix of form factors:
\beq
\label{eq:FF-matrix}
\mathcal{F}:=\begin{pmatrix}
\langle \bar{\psi} \bar{\psi} | \cO_F | 0 \rangle & \langle \bar{X} \bar{Y} \bar{Z} | \cO_F | 0 \rangle \\ \cr
\langle \bar{\psi} \bar{\psi} | \cO_B | 0 \rangle & \langle \bar{X} \bar{Y} \bar{Z} | \cO_B | 0 \rangle
\end{pmatrix}  \ . 
\eeq
Apart from the possible connections to phenomenologically relevant quantities in QCD alluded to earlier, there are additional reasons to study form factors of operators such as $\cO_B$ and $\cO_F$: 
\begin{itemize}
\item[{\bf 1.}]
Firstly, it is very interesting to scan the possible remainders of form factors of wider classes of non-protected operators, and compare to results obtained for protected operators and operators belonging to  different sectors. A key motivation is to search for regularities and determine universal building blocks in the results that are common to form factors of different operators. 
\item[{\bf 2.}] By computing loop corrections to minimal form factors of non-protected operators it is possible to find the dilatation operator. This was done recently at one loop for the complete one-loop dilatation operator in \cite{Wilhelm:2014qua} and at   two loops in the $SU(2)$ subsector \cite{Loebbert:2015ova}.
Potentially, this holds promise for gaining further insights into the integrability of $\cN=4$ SYM.%
\footnote{Complementary approaches based on two-point functions were recently explored in 
 \cite{Koster:2014fva,Brandhuber:2015boa,Brandhuber:2014pta}.}

\end{itemize}
The calculation of the two-loop remainder of the form factor $\langle \bar{X} \bar{Y} \bar{Z} | \cO_B | 0 \rangle$ is very instructive in this respect. Indeed, we will show that the remainder function is given by a sum of terms of decreasing transcendentality, where the leading, transcendentality-four term turns out to be identical to the remainder for the form factor $\langle \bar{X} \bar{X} \bar{X} | \Tr(X^3)  | 0 \rangle$  computed in \cite{Brandhuber:2014ica}.  Furthermore, the terms of transcendentality ranging from three to zero turn out to be related  to certain finite remainder densities introduced in  
\cite{Loebbert:2015ova} in the study of the dilatation operator in the $SU(2)$ sector. It is interesting that they appear (in some form) in the larger $SU(2|3)$ sector, possibly pointing to some universality of these quantities. 
This finding leads us to speculate that  the leading transcendental part of the correction terms to Higgs + multi-gluon processes induced by the interactions $O_i$, $i\geq1$, on the right-hand side of \eqref{corre}, can be equivalently obtained by computing their  form factors (or form factors related by supersymmetry) in the much simpler $\cN\!=\!4$ SYM theory. The fact that the maximally transcendental part  of the form factors in the $SU(2|3)$ sector is computed effectively by form  factors of half-BPS operators leads us to further speculate on the special role of such operators in computing the maximally transcendental part of the form factors of the operators $O_i$ for $i\geq1$ in QCD.

We will also study and resolve the operator mixing, a problem which requires the knowledge of the ultraviolet  (UV) divergences of  three additional form factors: $\langle \bar{X} \bar{Y} \bar{Z} | \cO_F | 0 \rangle$, 
$\langle \bar{\psi} \bar{\psi} | \cO_F | 0 \rangle$,  and $\langle \bar{\psi} \bar{\psi} | \cO_B | 0 \rangle$. Note that these four form factors are different in nature: while 
$\langle \bar{X} \bar{Y} \bar{Z} | \cO_B | 0 \rangle$ and 
$\langle \bar{\psi} \bar{\psi} | \cO_F | 0 \rangle$ are minimal (i.e.~the number of particles in the external state is the same as the number of  fields in the operator), $\langle \bar{\psi} \bar{\psi} | \cO_B | 0 \rangle$ is sub-minimal (more fields than particles), and $\langle \bar{X} \bar{Y} \bar{Z} | \cO_F | 0 \rangle$ is non-minimal. Furthermore, at the loop order we are working the latter two are free from infrared (IR) divergences, lacking a corresponding tree-level form factor.%
\footnote{We also note that  the discontinuities of sub-minimal form factors at two loops were computed in \cite{Zwiebel:2011bx} in complete generality.}
On the other hand they all have  UV divergences, which will be extracted to resolve the mixing and determine the two-loop dilatation operator in the $SU(2|3)$ sector, in agreement with \cite{Beisert:2003ys}. 
By diagonalising it, two distinguished combinations of $\cO_B$ and $\cO_F$ will be determined, one which is half-BPS \cite{Intriligator:1999ff, 
Bianchi:2001cm, Beisert:2003ys} and one which is a descendant of the Konishi operator 
\cite{Bianchi:2001cm, Beisert:2003ys, Eden:2005ve, Eden:2009hz}. 

The rest of the paper is organised as follows. In Sections \ref{sec2} and \ref{sec3} we will derive the form factor 
$\langle \bar{X} \bar{Y} \bar{Z} | \cO_B | 0 \rangle$
at one and two loops, respectively. The two-loop IR-finite (but still UV-divergent) remainder function is then derived in Section  \ref{sect:TwoLoopRemainder}. There we also establish relations of our result to the results of  \cite{Brandhuber:2014ica}  and 
\cite{Loebbert:2015ova} for the maximally and subleading transcendental pieces of our result, respectively.  In Section \ref{sect:NonMinimal} we compute the sub-minimal form factor $\langle \bar{X} \bar{Y} \bar{Z} | \cO_F | 0 \rangle$ 
 up to one loop, which is sufficient for the computation of the two-loop dilatation operator performed later. Section 
\ref{sec-subminimal} is devoted to computing   the sub-minimal form factor $\langle \bar{\psi} \bar{\psi} | \cO_B | 0 \rangle$ at two loops. Using the UV-divergent parts of these form factors, we compute in Section \ref{mixing} the two-loop dilatation operator in the $SU(2|3)$ sector, finding its eigenvectors and corresponding   anomalous dimensions up to two loops.  
% given in \eqref{twoldo}. 
We conclude with comments on potential future research directions in Section \ref{theend}. 
%\newpage

%%%%%%%%%%%%%%%%%%%%%%%%%%%%%%
%%%%%%%%%%%%%%%%%%%%%%%%%%%%%%

\section{One-loop minimal  form factor  $\langle\bar{X}\bar{Y}\bar{Z} |\Tr X[Y, Z] |0\rangle$}
\label{sec2}

In this section we consider form factors of the operator introduced in (\ref{eq:OB}),
\beq\nonumber
\O_B\, =\, \Tr(X[Y,Z])
\ , 
\eeq
at one loop. Before presenting  the calculation we summarise our notation and conventions for the reader's convenience.

\subsection{Setting up the  notation}

The fields appearing in the $SU(2|3)$ sector are 
\beq
\{X, Y, Z; \psi_{\alpha}\} 
\ , 
\eeq 
previously 
 introduced in \eqref{eq:XYZ-phi} and \eqref{eq:psi}.
We recall  that the fields $\phi_{AB}$ satisfy the reality condition
\beq
\label{reality} 
{\phi}^{AB}\ =\ \bar{\phi}_{AB}  \ =\  
{1\over 2}  \epsilon^{ABCD}\, \phi_{CD} \  , 
\eeq
and therefore 
\beq
\overline{X}\,=\,\phi_{34} = \phi^{12}\,,\quad \overline{Y}\,=\,\phi_{14}=\phi^{23}\,,\quad \overline{Z}\,=\,\phi_{24}=\phi^{31} \ .
\eeq
We also introduce
\beq
\psi_{ABC, \alpha}  = \epsilon_{ABCD} \, {\psi}^D_{\alpha} \ , \qquad \qquad 
\bar{\psi}^{ABC}_{\dot\alpha}  = \epsilon^{ABCD} \, \bar{\psi}_{D, \dot{\alpha}} \ . 
%= {1\over 3!} \epsilon_{ABCD} \psi^{BCD}\ , 
\eeq
In our conventions all on-shell particles appearing in amplitudes or form factors are
outgoing while the momentum $q$ of the off-shell operator in a form factor is by definition incoming. Therefore it is natural to introduce
the Nair super-annihilation operator  as 
\begin{align}
\begin{split}
\Phi (p, \eta)  & =   g^{(+)} (p) + \eta_A \psi^{A}(p)  + {1\over 2}  \phi^{AB}(p)
\eta_A \eta_B +   {1\over 3!} \bar{\psi}^{ABC}(p)  \eta_A \eta_B\eta_C 
 \\
 &+ g^{(-)}(p)  \eta_1\cdots \eta_4 \ , 
\end{split}
\end{align}
where $g^{(+)} (p)$,  $\psi^{A}(p), $  $\phi^{AB}(p)$, 
$\bar{\psi}^{ABC}(p)$ and  $g^{(-)}(p)$, denote the annihilation operators for the various particles of $\cN=4$ SYM.    
For instance $\langle 0 |\psi^{A}(p)$
is a state of an outgoing fermion with momentum $p$ and helicity $+1/2$, while $\langle 0 |\bar\psi^{ABC}(p) $ has momentum $p$ and helicity  $-1/2$. In the following we will usually denote multiparticle states with on-shell momenta 
$\langle \psi(p_i)^{A_i} \cdots \phi(p_j)^{A_j B_j} \cdots\bar\psi(p_k)^{A_k B_k C_k} \cdots |$ in the slightly more compact notation 
$\langle i^{\psi^{A_i}} \cdots j^{\phi^{A_j B_j}} \cdots k^{\bar\psi^{A_k B_k C_k}} \cdots |$ whenever we want to make particle labels explicit. Often we will also use the following shorthand notation if labels are not needed, in particular
$\langle \bar{X} \bar{Y} \bar{Z}| := \langle 1^{\phi^{12}} 2^{\phi^{23}} 3^{\phi^{31}}|$
and $\langle \bar \psi\bar \psi | :=\langle 1^{\bar{\psi}^{123}} 2^{\bar{\psi}^{123}} |$.

\subsection{A useful decomposition}

In order to compute  the form factor $\langle \bar{X} \bar{Y} \bar{Z} | \cO_B| 0\rangle$, with $\cO_B$ defined in \eqref{eq:OB}, 
we will make use of  the  decomposition
\begin{equation}
\label{eq:-2XZY}
\O_B\,=\,\tilde{\cO}_{\rm BPS}+\O_{\rm offset}\ ,
\end{equation}
where $\tilde{\cO}_{\rm BPS}$ is the half-BPS operator defined in \eqref{obps} and 
\beq
\label{offset}
\O_{\text{offset}}:=-2\,\Tr(XZY)
\ . 
\eeq
This decomposition turns out to be particularly useful for two reasons: 
\begin{itemize}
\item[{\bf 1.}] 
Firstly, it separates out the contribution of the half-BPS operator $\tilde{\O}_{\rm BPS}$. 
The result  for the corresponding half-BPS form factor  is identical to that of the half-BPS operator $\Tr(X^3)$ obtained  in  \cite{Penante:2014sza, Brandhuber:2014ica} up to two loops and need not be computed again.%
\footnote{See Appendix \ref{sect:compare-with-BPS} for details.} 
\item[{\bf 2.}]
Secondly,  the form factor  of the offset operator  $\langle \bar{X}\bar{Y}\bar{Z}  | \O_{\text{offset}} | 0\rangle$  turns out to be  particularly simple because of the ``shuffled" configuration of the state with respect to the fields inside the operator\footnote{Note that we could have performed the decomposition $\Tr(X[Y,Z]) =  -{\rm Tr}  (X  \{Y, Z\}) + 2 \, \Tr(XYZ)$ but this   is not  convenient for  our choice of external state $\langle \bar{X}\bar{Y}\bar{Z}|$.}.
 Specifically, we will find  that this form factor is expressed in terms of functions with strictly sub-maximal degree of transcendentality, while the half-BPS operator is expressed in terms of functions with maximal degree of transcendentality only.
 \end{itemize}
Therefore we focus on the ``offset" operator introduced in \eqref{offset}, 
from which the results for $\cO_B$ can then be easily obtained. 

\subsection{Two-particle cuts and result}
\label{oneloop}

\begin{figure}[h]
\centering
\includegraphics[width=0.35\linewidth]{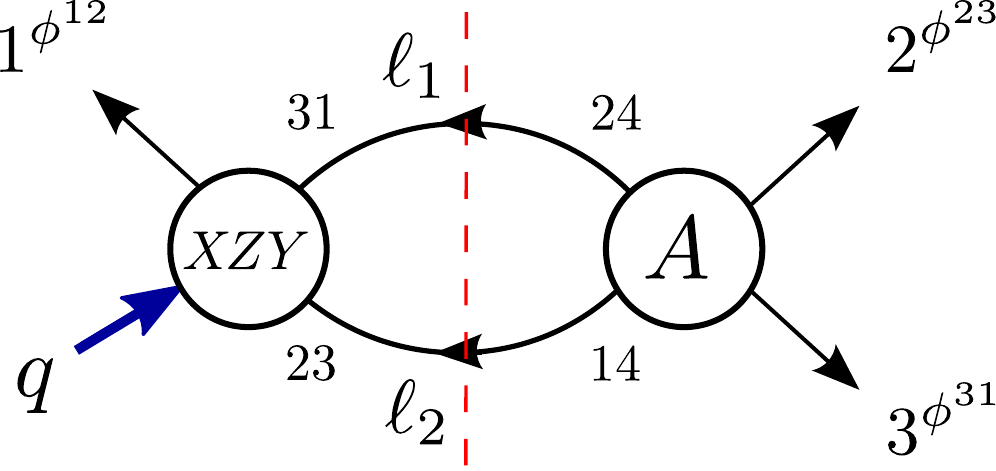}
\caption{\it Two-particle cut of the one-loop form factor $F^{(1)}_{\cO_{\rm offset}}(1^{\phi^{12}},2^{\phi^{23}},3^{\phi^{31}};q)$. We remind the reader of our notation: $X=\phi_{12}$, $Y=\phi_{23}$, $Z=\phi_{31}$, with 
$\bar{X}=\phi^{12}$, $\bar{Y}=\phi^{23}$ and $\bar{Z}=\phi^{31}$.}
\label{fig:XZY-to-XYZ-1-loop}
\end{figure}

In the following we denote by $F^{(L)}_{\cO_{\rm offset}}(1^{\phi^{12}},2^{\phi^{23}},3^{\phi^{31}};q)$ the $L$-loop contribution to the form factor  $\langle \bar{X}\bar{Y}\bar{Z}  | \O_{\text{offset}} (0)  | 0\rangle$. 
We begin by computing    $F^{(1)}_{\cO_{\rm offset}}(1^{\phi^{12}},2^{\phi^{23}},3^{\phi^{31}};q)$ with the  two-particle cut  shown in Figure \ref{fig:XZY-to-XYZ-1-loop}. This, plus two cyclic permutations of the external particles, are the only cuts contributing to this form factor. 

\noindent
The tree-level amplitude entering the cut is 
\begin{align}
&A(2^{\phi^{23}},3^{\phi^{31}},\ell_2^{\phi^{14}},\ell_1^{\phi^{24}})\,=\, i\, ,
\end{align}
while the required tree-level form factor is
\begin{align}
F^{(0)}_{\cO_{\rm offset}}(1^{\phi^{12}},-\ell_1^{\phi^{31}},-\ell_2^{\phi^{23}};q)=-2 \, .
\end{align}
Hence,  uplifting  the cut we simply get bubble integrals:%
\footnote{Note that each of the cut propagators carries an additional factor of $i$.} 
\begin{equation}
F^{(1)}_{\cO_{\rm offset}}(1^{\phi^{12}},2^{\phi^{23}},3^{\phi^{31}};q)\,=\, 2\,i\times\, \pic{0.4}{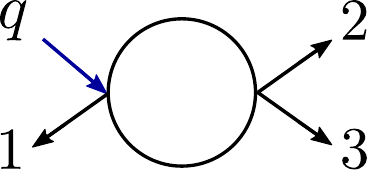}+\text{cyclic}(1,2,3)\,.
\end{equation}
% was - 2 i^3
A similar calculation shows that, as anticipated, 
 the one-loop form factor of the operator $\tilde{\cO}_{\rm BPS}$ introduced in \eqref{obps} is identical to that of the operator $\Tr(X^3)$ computed in 
 \cite{Penante:2014sza}, 
\begin{equation}
F_{\tilde{\cO}_{\rm BPS}}^{(1)}(1^{\phi^{12}},2^{\phi^{23}},3^{\phi^{31}};q)\,=\, i\, s_{23}\times\,\pic{0.4}{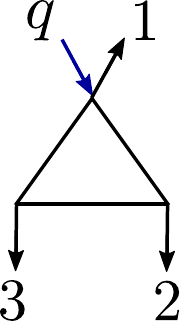}+\text{cyclic}(1,2,3)\ .
\end{equation}
% was -i^3
Thus, the one-loop form factor of $\cO_B$   is%
\footnote{Expressions for the one-loop master integrals can be found in Appendix \ref{App:Integrals}.}
\begin{equation}
\label{eq:XYZ-one-loop}
\,F_{\cO_B}^{(1)}(1^{\phi^{12}},2^{\phi^{23}},3^{\phi^{31}};q)\,=\, 2\,i\times\, \pic{0.4}{bubble-s23}+i\, s_{23}\times\,\pic{0.4}{triangle-s23}+\text{cyclic}(1,2,3)\,, 
\end{equation}
where  $s_{ij} := (p_i + p_j)^2$ as usual. 

From \eqref{eq:XYZ-one-loop} we can easily extract the one-loop anomalous dimension of $\cO_B$.
In order to extract the UV divergence from \eqref{eq:XYZ-one-loop} we have to remove the IR divergences which is achieved by simply dropping the triangle integrals. Using the results of Appendix \ref{App:Integrals}, we find
the UV divergence at the renormalisation scale $\mu_R$ to be
\beq
\left. F_{\cO_B}^{(1)}\right|_{\mu_R, {\mathrm{UV}}} \ = \ -{6 \over \epsilon} \,a (\mu_R)
\ , 
\eeq
where 
\beq
\label{Hooft-run}
a (\mu_R) \ :=\ {g^2 N  e^{- \epsilon \gamma_{\rm E}}\over  (4 \pi )^{2-\epsilon }}\left( {\mu_R \over \mu}\right)^{- 2 \epsilon} 
\ , 
\eeq 
and  $\mu$ is the usual dimensional regularisation mass parameter. 
From this we can read off the  one-loop anomalous dimension via 
\beq
\gamma_{\cO}  = - \mu_R {\partial \over \partial \mu_R} \left.\log  (1 + \cZ^{(1)}_{\cO} + \cdots )\right|_{\eps \to 0} 
\ ,
\eeq 
with 
\beq
\cZ^{(1)}_{\cO_B} \ = \ {6\over \epsilon} \ a (\mu_R)  
\ . 
\eeq
This leads to
 \beq
 \label{eq:one-loop-anomalous-dim}
 \gamma^{(1)}_{\cO_B} = 12 \, a
 \ ,
 \eeq
 where $a$ is the four-dimensional 't Hooft coupling, given by
 \beq 
 \label{Hooft}
a :=  {g^2 N \over  ( 4 \pi)^2}\ .
\eeq 
The result \eqref{eq:one-loop-anomalous-dim} is in agreement with known results for the one-loop anomalous dimension of the Konishi multiplet.
The same value can be obtained with an explicit application of the  formula  for the complete one-loop  dilatation operator of  \cite{Zwiebel:2011bx}.

\subsection{Auxiliary one-loop form factors needed for two-loop cuts}
\label{additional}

In this section we discuss two additional one-loop form factors that will appear as building blocks for the two-particle cuts of the two-loop form factor of $\cO_{\rm offset}$ (and  thus $\cO_B$) in   Section \ref{aux}.

\begin{figure}[h]
\centering
\includegraphics[width=0.35\linewidth]{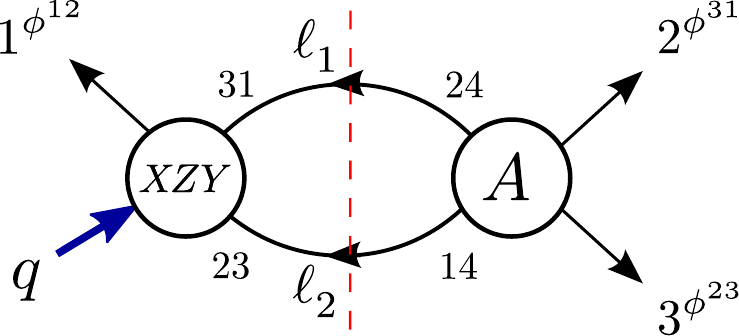}
\caption{\it One of the three  two-particle cuts of the one-loop form factor $F^{(1)}_{\cO_{\rm offset}}(1^{\phi^{12}},2^{\phi^{31}},3^{\phi^{23}};q)$. Two more cuts are obtained by cyclically permuting the external legs.
}
\label{fig:XZY-to-XYZ-1-loop-bis}
\end{figure}

The first form factor we   consider  is   $F^{(1)}_{\cO_{\rm offset}}(1^{\phi^{12}},2^{\phi^{31}},3^{\phi^{23}};q)$, where now the ordering of the particles in the state parallels that of the fields in the operator. A simple two-particle cut is sufficient to determine it,  see  Figure \ref{fig:XZY-to-XYZ-1-loop-bis}.
%\\

\begin{figure}[h]
\centering
\includegraphics[width=0.8\linewidth]{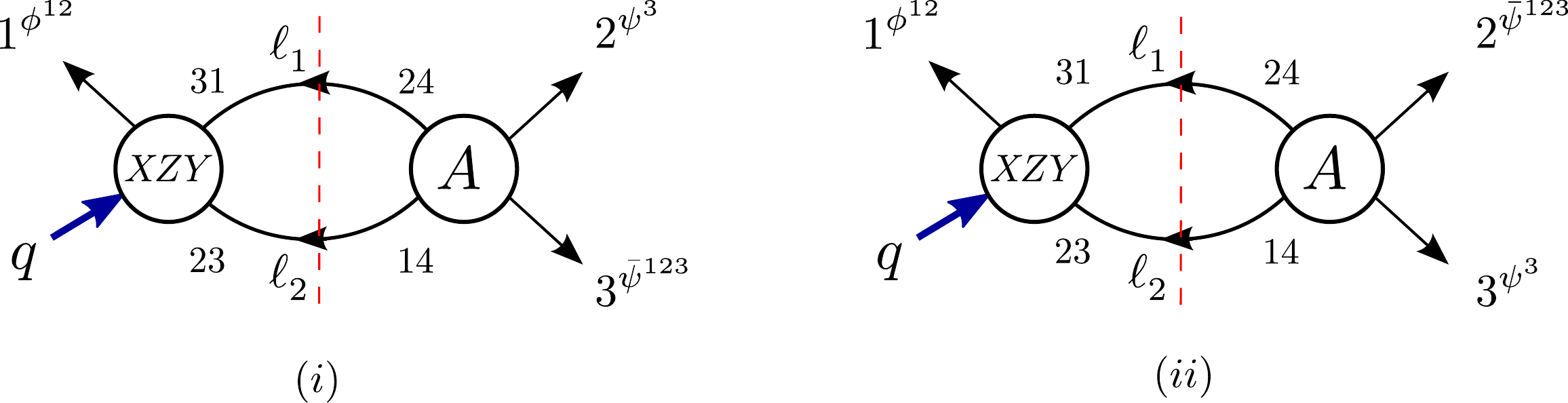}
\caption{\it One-loop form factors with a fermionic external state  entering the two-loop two-particle cuts of  Figure \ref{fig:two-loop-fermions}.}
\label{fig:one-loop-fermions}
\end{figure}

\noindent
The amplitude entering the cut is 
\beq A(2^{\phi^{31}},3^{\phi^{23}},\ell_2^{\phi^{14}},\ell_1^{\phi^{24}})\,=\, i\,\frac{\b{2 \ell_2}\b{3 \ell_1}}{\b{3 \ell_2}\b{\ell_1 2}}\ ,  
\eeq
thus we get 
\beq 
F^{(1)}_{\cO_{\rm offset}}(1^{\phi^{12}},2^{\phi^{31}},3^{\phi^{23}};q)\,=\, -2\,i\times\pic{0.4}{bubble-s23}-2\,i\,s_{23}\times 
\pic{0.4}{triangle-s23}+\text{cyclic}(1,2,3)\ .
\eeq
Next we consider the form factors of $\O_{\rm offset}$ with a fermionic external state made of excitations $\psi^3$ and $\bar{\psi}^{123}$, as shown in 
Figure~\ref{fig:one-loop-fermions}. 

The results for the two-particle cuts for the two independent orderings of the fermionic  legs are
\begin{align}
\begin{split}
\label{eq:one-loop-fermions}
(i):\ & F^{(1)}_{\cO_{\rm offset}}(1^{\phi^{12}},2^{\psi^{3}},3^{\bar{\psi}^{123}};q)\Big|_{2,s_{23}}\,=\,-2\,i^2 A(2^{\psi^3},3^{\bar{\psi}^{123}},\ell_2^{\phi^{14}},\ell_1^{\phi^{24}})\,=\,-2i\,[2 |\ell_1| 3\rangle \times\pic{0.4}{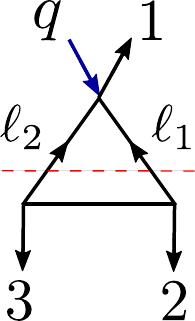}  \ ,\\[5pt]
(ii):\ & F^{(1)}_{\cO_{\rm offset}}(1^{\phi^{12}},2^{\bar{\psi}^{123}},3^{\psi^{3}};q)\Big|_{2,s_{23}} \,=\,  -2\,i^2  A(2^{\bar{\psi}^{123}},3^{\psi^{3}},\ell_2^{\phi^{14}},\ell_1^{\phi^{24}})\,=\,-2i\,\langle2|\ell_2|3]\times \pic{0.4}{triangle-s23-ell}  \ ,
\end{split}
\end{align}
where we denote the $m$-particle cut of an $L$-loop form factor of an operator $\O$ in a generic $P^2$-channel by 
\begin{equation}
F^{(L)}_{\O} (\dots;q)\Big|_{m,P^2}\ .
\end{equation}
Both form factors are expressed in terms of a linear triangle which we refrain from reducing to scalar integrals since we are working at the integrand level.%
\footnote{Furthermore,  both  expressions would vanish upon performing the loop integration. 
Indeed  by Lorentz invariance, after Passarino-Veltman reduction  one would have e.g.~for the first form factor $\ell_1 \to a p_2 + b p_3$, thus $[2|\ell_1 | 3\rangle \to 0$ after the reduction.}
Instead we will plug these expressions into the two-particle cuts of the two-loop form factors shown in Figure \ref{fig:two-loop-fermions}.

\section{Two-loop minimal  form factor  $\langle\bar{X}\bar{Y}\bar{Z} |\Tr X[Y, Z] |0\rangle$}
\label{sec3}

We proceed to compute  the minimal form factor of $\cO_B=\Tr(X[Y,Z])$ at two loops with the external state $\langle \bar{X}\bar{Y}\bar{Z} |$.   The strategy of the calculation is as follows:
\begin{itemize}
\item[{\bf 1.}]
Thanks to the decomposition  \eqref{eq:-2XZY}, we  need only  compute   the form factor of the operator $\O_{\text{offset}}\!=\!-2\,\Tr(XZY)$. This will be done in Sections \ref{sec:two-loop-two-particle-cuts} and  \ref{sec:two-loop-three-particle-cuts}.
\item[{\bf 2.}]
We then obtain the required form factor of $\cO_B$ by adding to our result that of the half-BPS operator $\tilde{\cO}_{\rm BPS}\!=\!\Tr ( X\{Y, Z\})$, which is  identical to the form factor $\langle  \bar{X}\bar{X}\bar{X} | \Tr (X^3) |0 \rangle$ computed in \cite{Brandhuber:2014ica}, which we quote here for the reader's convenience:
\begin{align}\label{BPS}
\hspace{-1cm}\begin{split}
F_{\tilde{\cO}_{\rm BPS}}^{(2)}\,=\,&-\sum_{i=1}^3\pic{0.35}{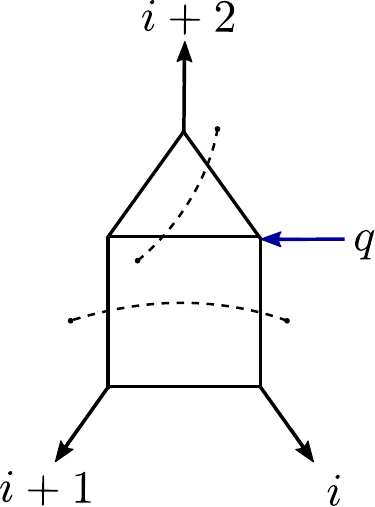} \ + \pic{0.35}{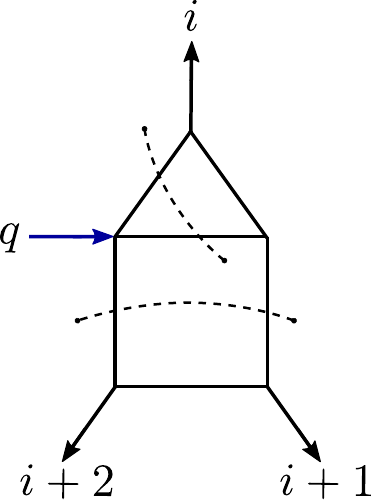} \ + \pic{0.35}{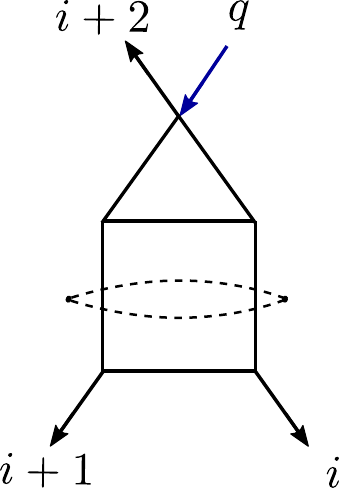}
 + \pic{0.35}{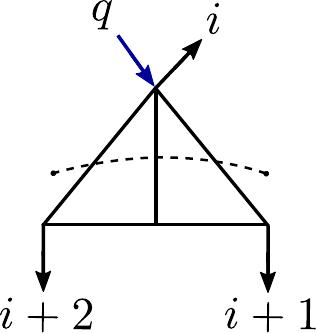} \ -\pic{0.35}{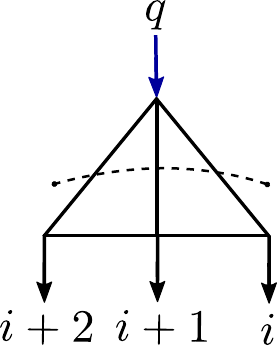}\ . 
\end{split}
\end{align}
In order to define the numerators we use the notation introduced in {\rm \cite{Brandhuber:2014ica}}: each dashed line corresponds to a numerator factor equal to the  total momentum flowing through it, squared. For example, the third integral in \eqref{BPS} comes with the 
factor $(s_{i\,i+1})^2$.
\item[{\bf 3.}]
In Section \ref{sect:summary-and-reduction} we summarise the complete result and perform the integral reduction.
\end{itemize}

\subsection{Two-particle cuts of the two-loop form factor}
We begin by considering the possible two-particle cuts of the two-loop form factor. There are two types of cuts to consider, which are of the form $F^{(0)}\times A^{(1)}$ and $F^{(1)}\times A^{(0)}$.

\label{sec:two-loop-two-particle-cuts}

\subsubsection{Tree-level form factor $\times$ one-loop amplitude}

The first two-particle cut we consider is of the form $F^{(0)}\times A^{(1)}$, and we will focus on the  $s_{23}$-channel. The other cuts are obtained by cyclically permuting the external legs.

\begin{figure}[h]
\centering
\includegraphics[width=0.35\linewidth]{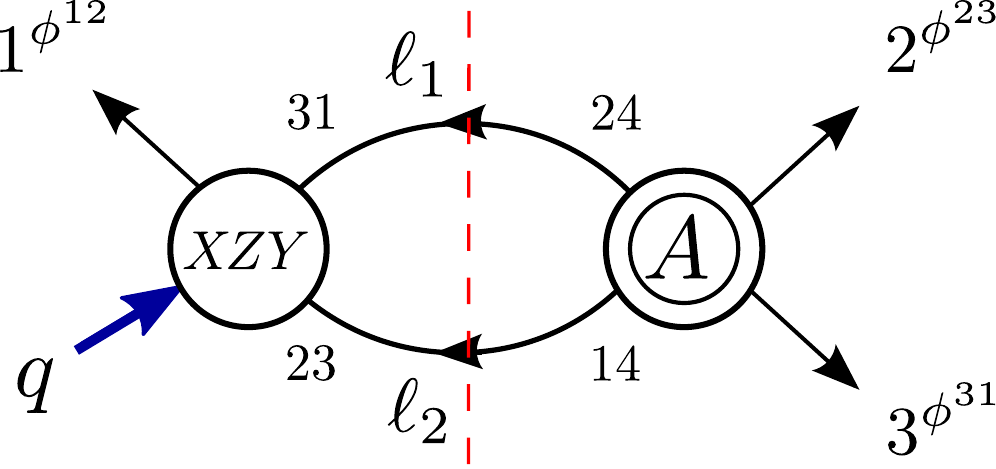}
\caption{\it Two-particle cut contributing to the two-loop form factor in the $s_{23}$-channel.}
\end{figure}
In this case the one-loop amplitude is
\beq
A^{(1)}\,=\, A^{(0)}\Big[-s_{12}s_{23}\times\pic{0.4}{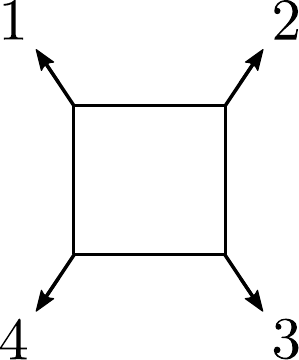}\Big] \ ,
\eeq
hence the  algebra of the previous section iterates and we~get the following result for the~cut:
\begin{align}
\begin{split}
&F^{(2)}_{\cO_{\rm offset}}(1^{\phi^{12}},2^{\phi^{23}},3^{\phi^{31}};q)\Big|_{2,s_{23}}\,=\,\qquad -2\,s_{23}s_{2\ell_1}\times\pic{0.55}{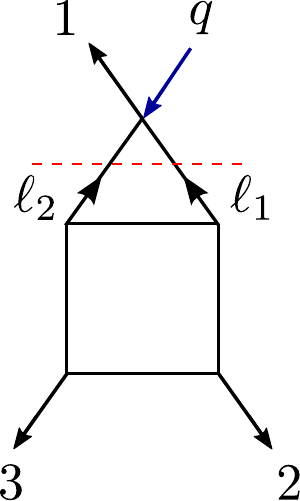}\, .
\end{split}
\end{align}

\subsubsection{One-loop form factor $\times$ tree-level amplitude}
\label{aux}

\begin{figure}[h]
\centering
\includegraphics[width=0.8\linewidth]{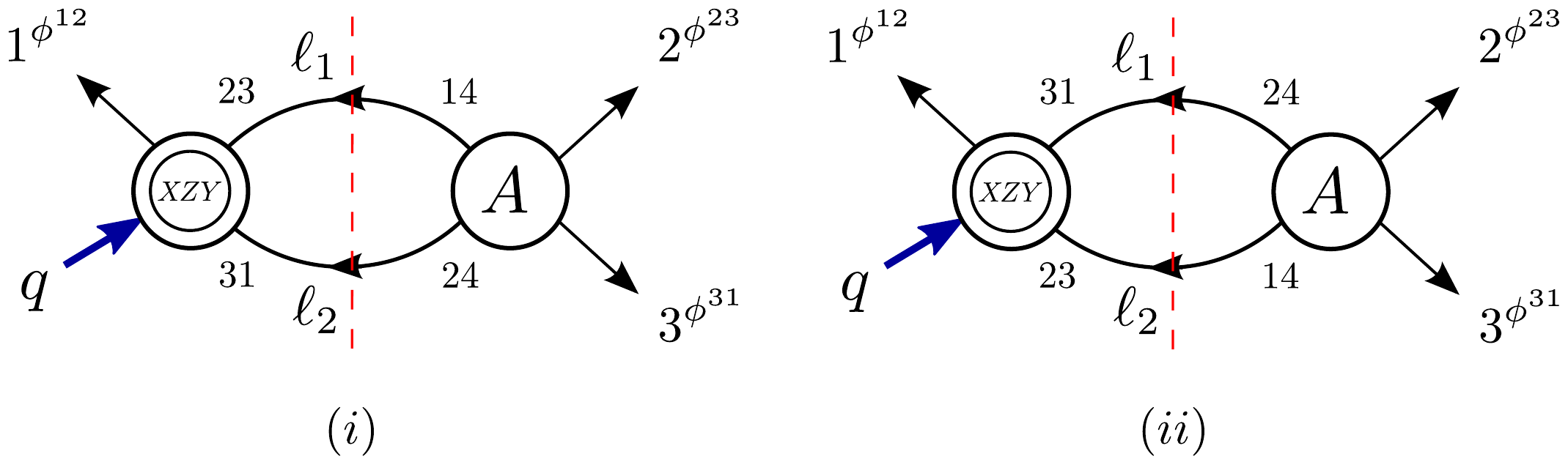}
\caption{\it Contribution to the two-loop form factor from  scalars  in the loop.}
\label{fig:two-loop-scalars}
\end{figure}

\begin{figure}[h]
\centering
\includegraphics[width=0.75\linewidth]{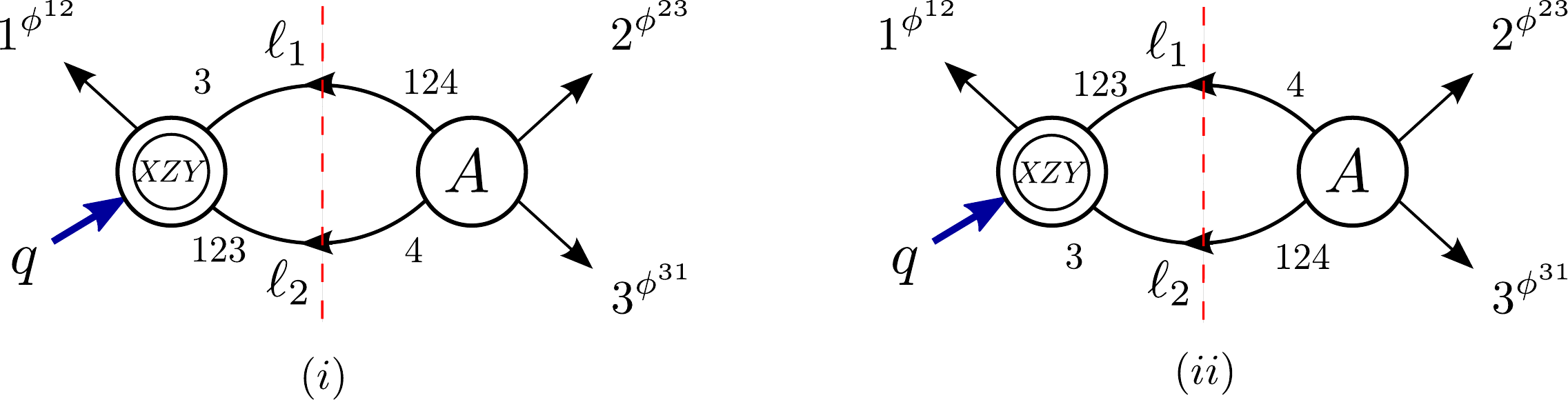}
\caption{\it Two-loop form factors with internal fermions. The one-loop form factors on the left-hand-side of the cuts were computed in \eqref{eq:one-loop-fermions}.}
\label{fig:two-loop-fermions}
\end{figure}

Next we consider two-particle cuts of the form $F^{(1)}\times A^{(0)}$.
There are  two options for the states running in the loop:  we can either have  scalars, as shown in Figure~\ref{fig:two-loop-scalars}, or fermions, as in Figure~\ref{fig:two-loop-fermions}. We consider these two types of contributions in turn. 

%%%%%%%%%%%%%%

\noindent
{\bf Scalars in the loop}. This case is illustrated in Figure \ref{fig:two-loop-scalars}. 
The relevant one-loop form factors were calculated in Section \ref{oneloop}, while the tree amplitudes entering the cuts are \begin{align}
\begin{split}
(i):\quad&\,\,\,A(2^{\phi^{23}},3^{\phi^{31}},\ell_2^{\phi^{24}},\ell_1^{\phi^{14}})\,=\,  i\frac{\b{2 \ell_2}\b{3 \ell_1}}{\b{3 \ell_2}\b{\ell_1 2}}\,=\,-i\left(1 + \frac{s_{23}}{2(\ell_1\cdot p_2)}\right) \,,\\
(ii):\quad&\,A(2^{\phi^{23}},3^{\phi^{31}},\ell_2^{\phi^{14}},\ell_1^{\phi^{24}})\,=\,i\,.
\end{split}
\end{align}
This results in the following  possibilities:
\begin{align}
\label{eq:two-loop-q-p1}
\begin{split}
&F^{(2)}_{\cO_{\rm offset}}(1^{\phi^{12}},2^{\phi^{23}},3^{\phi^{31}};q)\Big|^{\rm scalars}_{2,s_{23}}\,=\, -4 \times \Bigg[\; \pic{0.4}{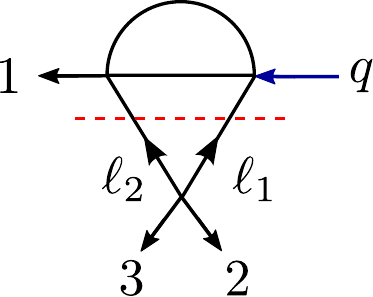} + \pic{0.4}{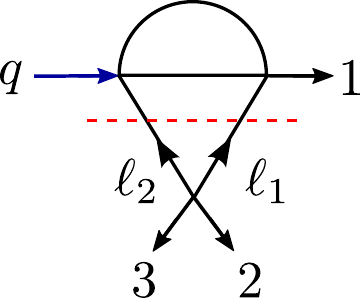} \\
& +\pic{0.4}{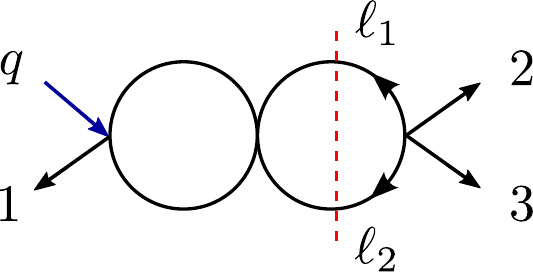}\quad \Bigg] -2 \times \Bigg[  s_{1\ell_2}\times \pic{0.4}{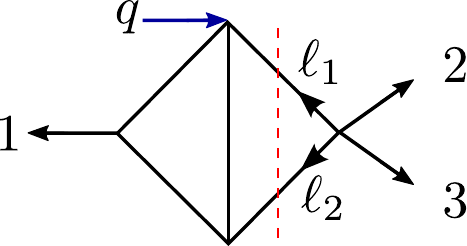}  +s_{1\ell_1}\times\pic{0.4}{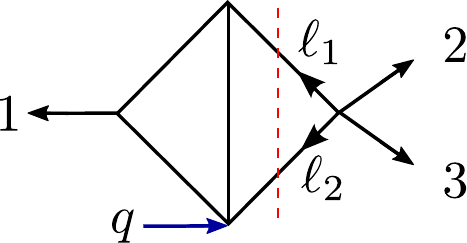}\quad \Bigg]  \\[15pt]
&  - 2\,s_{23} \times \Bigg[\quad \pic{0.4}{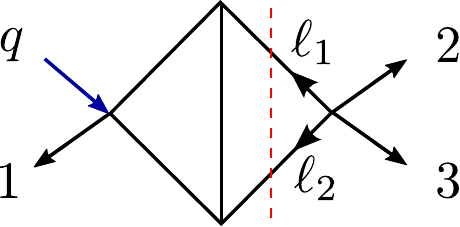} + \pic{0.4}{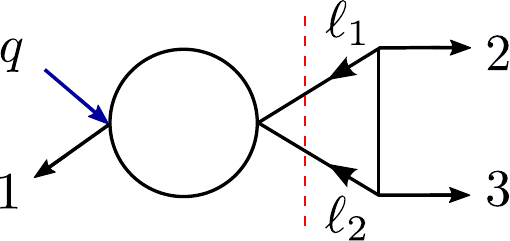} + \pic{0.4}{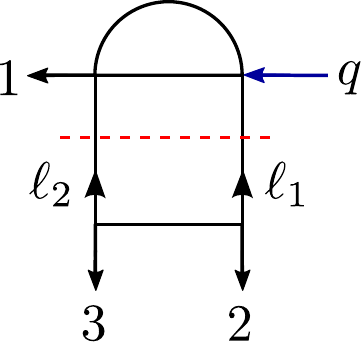} + \pic{0.4}{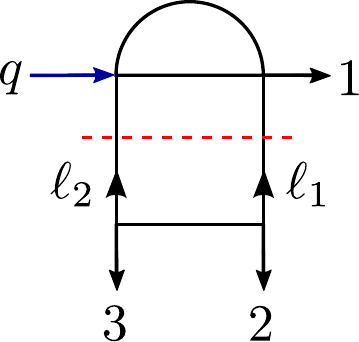} \quad \Bigg]\,.\\
\end{split}
\end{align}
Note that all the topologies which have a one-loop sub-amplitude containing triangles or bubbles have to cancel  as a consequence of the amplitude no-triangle theorem  \cite{Bern:1994zx}  --  these are the integrals number 3, 6 and 7 in \eqref{eq:two-loop-q-p1}. This cancellation  occurs after adding the contribution from fermions running in the loop, which we compute now. 

%%%%%%%%%%%%%%

\noindent
{\bf Fermions in the loop.}
The contribution  from  fermions  in the loop  are shown in Figure~\ref{fig:two-loop-fermions}.

We use the expressions for the one-loop form factors given in \eqref{eq:one-loop-fermions}\footnote{We added an extra minus sign to every expression to take into account the reversal of direction of $\ell_1$ and $\ell_2$ according to the following practical prescription \cite{Carrasco:2015iwa}: $\lambda_{-P}=-\lambda_P$, $\tilde\lambda_{-P}=\tilde\lambda_P$, $\eta_{-P}=\eta_P$. } and amplitudes, graphically represented as:
\begin{align}
& F^{(1)}_{\cO_{\rm offset}}(1^{\phi^{12}},-\ell_1^{\psi^{3}},-\ell_2^{\bar{\psi}^{123}};q)\,=\, 2i [\ell_1|\ell_4|\ell_2\rangle \times \pic{0.4}{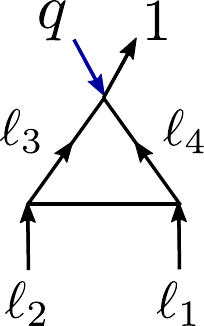}\ ,\\[5pt]
& F^{(1)}_{\cO_{\rm offset}}(1^{\phi^{12}},-\ell_1^{\bar{\psi}^{123}},-\ell_2^{\psi^{3}};q)\,=\, 2i \langle\ell_1|\ell_3|\ell_2] \times \pic{0.4}{triangle-s23-ell1234}\ ,
\end{align}
\begin{align}
&A(2^{\phi^{23}},3^{\phi^{31}},\ell_2^{\psi^{4}},\ell_1^{\bar\psi^{124}})\,=\,-i[\ell_2 |3 |\ell_1\rangle \times \pic{0.4}{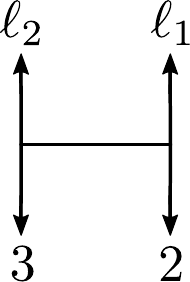}\ ,\\
&A(2^{\phi^{23}},3^{\phi^{31}},\ell_2^{\bar\psi^{124}},\ell_1^{\psi^{4}})\,=\,-i\langle\ell_2 |2 |\ell_1] \times \pic{0.4}{tree-amp-fermion}\ .
\end{align}

\noindent We obtain the following results for the cuts shown in Figure \ref{fig:two-loop-fermions}:
\begin{align}
\begin{split}
\label{eq:fermion-two-loop-cut1}
(i):\quad &-i^2\,F^{(1)}_{\cO_{\rm offset}}(1^{\phi^{12}},-\ell_1^{\psi^{3}},-\ell_2^{\bar{\psi}^{123}};q)\times A(2^{\phi^{23}},3^{\phi^{31}},\ell_2^{\psi^{4}},\ell_1^{\bar\psi^{124}})\\ 
\,=\,& 2 \, [\ell_1 |\ell_4| \ell_2\rangle[\ell_2| 3| \ell_1\rangle \times \pic{0.4}{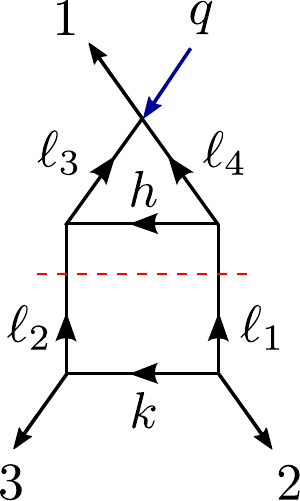}\,
\end{split}\\[15pt]
\begin{split}
\label{eq:fermion-two-loop-cut2}
(ii):\quad & -i^2\,F^{(1)}_{\cO_{\rm offset}}(1^{\phi^{12}},-\ell_1^{\bar{\psi}^{123}},-\ell_2^{\psi^{3}};q)\times A(2^{\phi^{23}},3^{\phi^{31}},\ell_2^{\bar\psi^{124}},\ell_1^{\psi^{4}})\\ 
\,=\,& 2\,\langle\ell_1|\ell_3|\ell_2]\langle \ell_2| 2| \ell_1]\times \pic{0.4}{SOTT-cut123-fermions-master} \ ,
\end{split}
\end{align}
where for convenience we have labeled the additional internal momenta  as $k$ and $h$, 
%\red 
and we have also multiplied the result of the cut by $(-1)$ from the fermion loop. Note that  $\ell_1$ and $\ell_2$ are cut, while $\ell_3,\,\ell_4,\,k$ and $h$ are  off shell.\\

\noindent 
Combining  \eqref{eq:fermion-two-loop-cut1} and  \eqref{eq:fermion-two-loop-cut2}  we obtain
\begin{align}
\label{eq:traces-fermions}
\begin{split}
%-2\,XZY:\quad 
&\Big[ \eqref{eq:fermion-two-loop-cut1}  + \eqref{eq:fermion-two-loop-cut2}\Big] \,=\,2\Big[\Tr_+(2\,\ell_1 \ell_4 \ell_2)+\Tr_+(2\,\ell_2 \ell_4 \ell_1)\Big]\times \pic{0.4}{SOTT-cut123-fermions-master}\ ,
\end{split}
\end{align}
where we used momentum conservation $\ell_1+\ell_2=\ell_3+\ell_4=-p_2-p_3$ and the fact that on the cut $s_{2\ell_1}=s_{3\ell_2}$.
Next we evaluate the traces in  \eqref{eq:traces-fermions} 
and 
expand the  various scalar products in  terms of the  inverse  propagators appearing in the main topology above, specifically using
\begin{align}
\begin{split}
2(\ell_2\cdot \ell_3)\,&=\,2(\ell_1\cdot\ell_4)+\ell_3^2-\ell_4^2\,=\, -h^2+\ell_3^2\ ,\\
2(\ell_4\cdot \ell_2)\,
&=\, s_{23} +h^2 -\ell_3^2 \ ,\\
2(p_2\cdot\ell_2)\,&=\, -2(p_2\cdot\ell_1) -s_{23}\,=\,-k^2-s_{23}\ ,
\end{split}
\end{align}
where $k^2=(p_2+\ell_1)^2\,=\,2(p_2 \cdot \ell_1)$ and $h^2=(\ell_1-\ell_4)^2\,=\,-2(\ell_1 \cdot \ell_4)+\ell_4^2$.
Doing so, we can  rewrite  \eqref{eq:traces-fermions} and obtain  the fermionic contribution to the two-particle cut of the two-loop form factors of  $\O_{\rm offset}$, 
\begin{align}
\begin{split}
\hspace{-0.5cm}
\label{eq:fermion-cut}
& F_{\cO_{\rm offset}}^{(2)}(1^{\phi^{12}},2^{\phi^{23}},3^{\phi^{31}};q)\Big|^{\rm fermions}_{2,s_{23}}=\, 2\Big[2\, k^2\, h^2 + s_{23}(k^2+h^2) - k^2(\ell_3^2+\ell_4^2)-s_{23}s_{2\ell_4}\Big]\times  \pic{0.4}{SOTT-cut123-fermions-master} .
\end{split}
\end{align}
From \eqref{eq:fermion-cut} we can now proceed to work out  the cut integrals contributing to the form factor of $\O_{\rm offset}$. 
We arrive at the result 
\begin{align}
\label{eq:fermionic-topology-result}
\begin{split}
&F^{(2)}_{\cO_{\rm offset}}(1^{\phi^{12}},2^{\phi^{23}},3^{\phi^{31}};q)\Big|^{\rm fermions}_{2,s_{23}}\,=\, 
 2s_{23}\times\Bigg[\;\pic{0.4}{kite2-cut123} + \pic{0.4}{bubble-triangle-cut123}\Bigg]\\
&- 2\times\Bigg[\;s_{23}s_{3\ell} \times \pic{0.4}{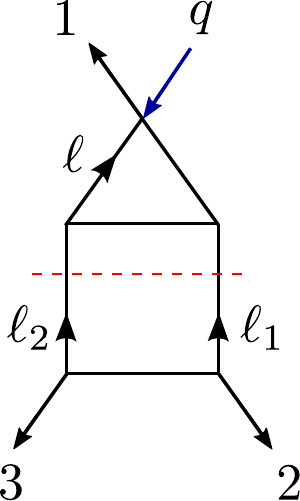} + \pic{0.4}{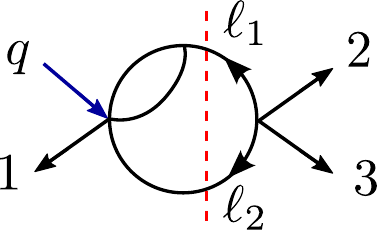}+\pic{0.4}{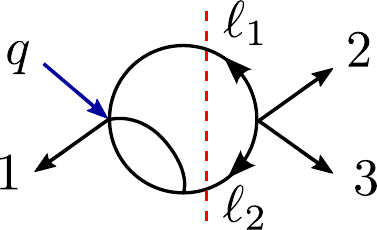}-2\times \pic{0.4}{double-bubble-cut123}\;\Bigg]\ .
\end{split}
\end{align}
We observe that the first, second and last integral in \eqref{eq:fermionic-topology-result}  precisely cancel the unwanted contributions in \eqref{eq:two-loop-q-p1}. 

%%%%%%%%%%%%%

\subsubsection{Result of two-particle cuts}
It remains to sum up the scalar and fermion contributions to the cut in question, given in \eqref{eq:two-loop-q-p1} and \eqref{eq:fermionic-topology-result}, respectively. The combined result is:
\begin{align}
\begin{split}\label{summary-2pcut}
&F^{(2)}_{\cO_{\rm offset}}(1^{\phi^{12}},2^{\phi^{23}},3^{\phi^{31}};q)\Big|_{2,s_{23}}\,=\,-2s_{23}s_{3\ell} \times \pic{0.4}{SOTT-cut123-result} +2s_{1\ell_2}\times \pic{0.4}{kite-up-cut123}\\
& +2s_{1\ell_1}\times\pic{0.4}{kite-down-cut123}-2s_{23}\times\Bigg[\quad \pic{0.4}{R2D2right-cut123} + \pic{0.4}{R2D2left-cut123}\quad\Bigg]\\
& - 2\times\Bigg[\quad \pic{0.4}{teardrop-up-cut123}+\pic{0.4}{teardrop-down-cut123}\quad\Bigg] -4\times\Bigg[\quad\pic{0.4}{ice-cream-left-cut123} +\pic{0.4}{ice-cream-right-cut123}\quad\Bigg]\ .
\end{split}
\end{align}
Note that the unwanted topologies which would lead to a violation of the amplitude no-triangle theorem  \cite{Bern:1994zx} have cancelled, as expected. 
We also observe that some of the numerators in \eqref{summary-2pcut} are ambiguous due to the cut conditions, and will be determined from three-particle cuts. 

%%%%%%%%%%%%%

\subsection{Three-particle cuts of the two-loop form factor} \label{sec:two-loop-three-particle-cuts}
In this section we study the three-particle cuts of the form factor of the operator  $\cO_{\rm offset}$ defined in \eqref{offset} at two loops. 
This computation will allow us to fix ambiguities of the numerators of integrals obtained from two-particle cuts and, in addition, provide additional integrals which are not detected by two-particle cuts.  
We consider three-particle cuts in the   $q^2$-channel  in Section \ref{sec:3-part-cut-q},  and  in the $s_{23}$-channel in Section \ref{sec:3-part-cut-q-p1}.

%%%%%%%%%%%%%

\subsubsection{Three-particle cuts in the $q^2$-channel}
\label{sec:3-part-cut-q}
\begin{figure}[h]
\centering
\includegraphics[width=1\linewidth]{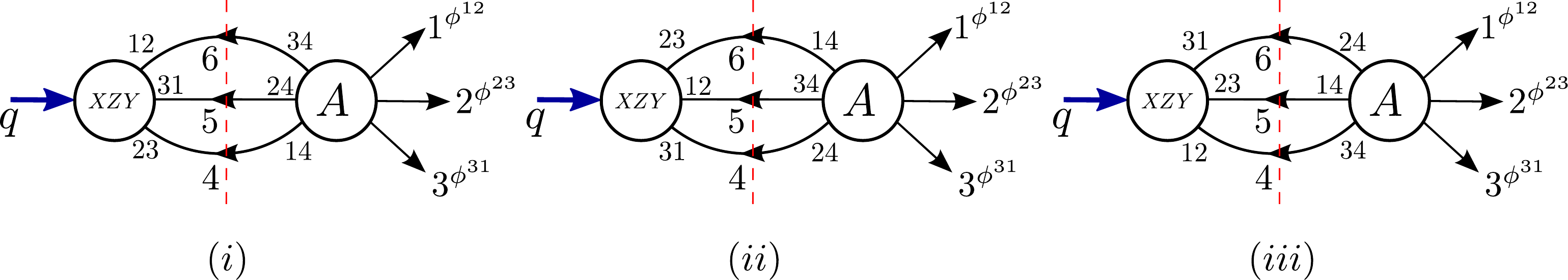}
\caption{\it Three-particle cuts in the the $q^2$-channel.}
\label{fig:q2-triple-cut}
\end{figure}

We begin by studying  the three independent  $q^2$-channel cuts shown in Figure \ref{fig:q2-triple-cut}.

\noindent
The corresponding    six-point scalar amplitudes are:
\begin{align}
\begin{split}
&A(1^{\phi^{12}},2^{\phi^{23}},3^{\phi^{31}},4^{\phi^{14}},5^{\phi^{24}},6^{\phi^{34}})\,=\,  
i \Big[ 
\frac{1}{s_{126}} +\frac{1}{s_{234}}-\frac{1}{s_{16}}  +\frac{s_{12}}{s_{16}s_{126}} +\frac{s_{56}}{s_{16}s_{234}}
\Big] 
 \,, \\
 &A(1^{\phi^{12}},2^{\phi^{23}},3^{\phi^{31}},4^{\phi^{24}},5^{\phi^{34}},6^{\phi^{14}})\,=\, i \Big[ 
 \frac{1}{s_{126}}+ \frac{1}{s_{234}} -\frac{1}{s_{34}}  +\frac{s_{23}}{s_{34}s_{234}} +\frac{s_{45}}{s_{34}s_{126}}\Big] 
\,, \\
&A(1^{\phi^{12}},2^{\phi^{23}},3^{\phi^{31}},4^{\phi^{34}},5^{\phi^{14}},6^{\phi^{24}})\,=\, 0\, ,
\end{split}
\end{align}
where to simplify the notation we  have  called the cut legs $p_4$, $p_5$ and $p_6$.
We can now immediately read off the contributions to the three-particle cuts:%
\footnote{Recall that  $\O_{\rm offset} = -2\Tr ( XZY)$, and note that in the cuts  the factor of $i$ from the amplitude cancels with the factor of $i^3$ from the  three   propagators.}
\begin{align}
\begin{split}
\label{3pc}
&F^{(2)}_{\cO_{\rm offset}}(1^{\phi^{12}},2^{\phi^{23}},3^{\phi^{31}};q)\Big|_{3,q^2}\,=\, 
-4\times \Bigg[  \pic{0.4}{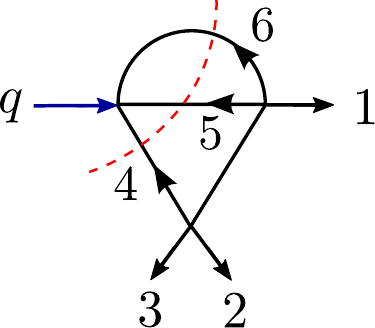}+\pic{0.4}{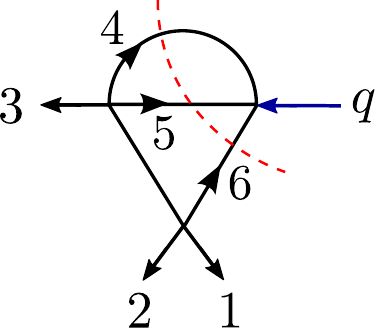}\Bigg] \\
&+ 2\times \Bigg[\quad \pic{0.4}{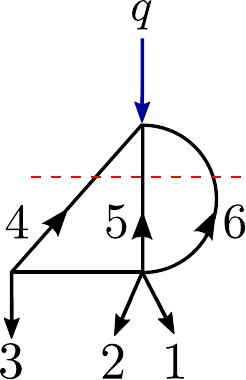}\quad+\quad\pic{0.4}{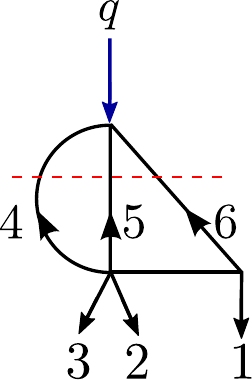}\quad \Bigg]  - 2 s_{12}\times \pic{0.4}{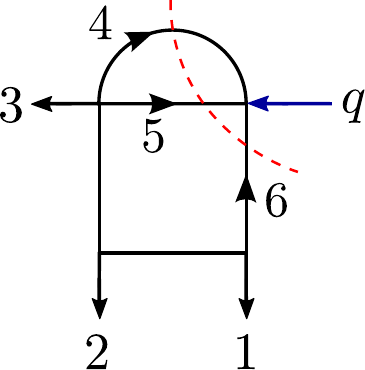}-2s_{23}\times \pic{0.4}{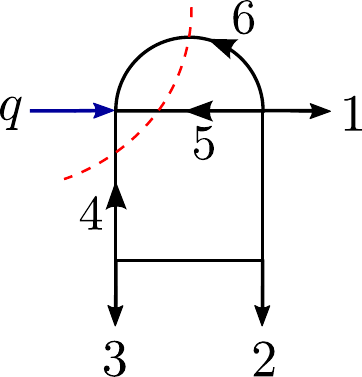}\\
&-2s_{1\ell}\times \pic{0.4}{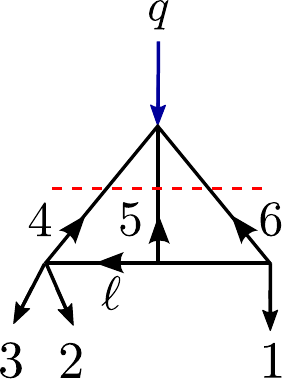}\quad-2s_{3\ell}\times \pic{0.4}{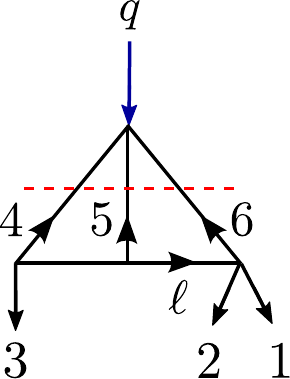}\ .
\end{split}
\end{align}
Two observations are in order. Firstly, new  topologies have appeared, which do not have two-particle cuts. 
Furthermore, the ambiguities we had found in some of the numerators of topologies identified using two-particle cuts have now been resolved.

As a final set of consistency checks, we now perform additional three-particle cuts in the $s_{23}$-channel. 

%%%%%%%%%%%%%%%%%%%%%%%

\subsubsection{Three-particle cuts in the $s_{23}$-channel}
\label{sec:3-part-cut-q-p1}

In this cut, $R$-symmetry allows for two possibilities for the particles running in the loop, namely   two scalars and a gluon, or two fermions and a scalar. 
There are two distinct situations  to consider, namely 
\beq
{F}^{\rm \overline{MHV}} \times {A}^{{\rm{MHV}}}
\qquad {\rm and} \qquad 
{F}^{\rm MHV} \times {A}^{\overline{\rm{MHV}}}
\ . 
\eeq
We now  study the first case in detail, while the second can be obtained by just interchanging $\la\cdot ,\cdot \ra \leftrightarrow \ls\cdot ,\cdot \rs$ and simply doubles up the contribution from the first case.
As before, we focus our attention on the  operator $\O_{\rm offset}$ introduced in~\eqref{offset}.\\

\begin{figure}[h]
\centering
\includegraphics[width=\linewidth]{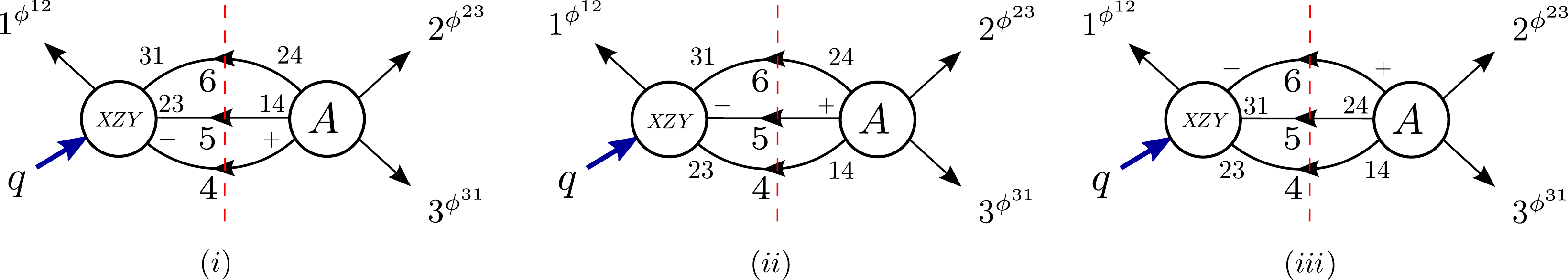}
\caption{\it Three cut diagrams  for the case of  a single gluon running in one of the internal loop legs. There are three more diagrams where the internal gluon has the opposite helicity. These are obtained by parity conjugation of the diagrams in this figure. }
\label{fig:XZY-q+1-gluons-first}
\end{figure}

\noindent
{\bf Gluons in the loop.}
The gluon can be exchanged in any of the three loop legs,  as shown in Figure \ref{fig:XZY-q+1-gluons-first}.

\noindent
 The corresponding integrands are
\begin{align}
%\begin{split}
(i):& \,i^3\, A(2^{\phi^{23}},3^{\phi^{31}}, 4^{+}, 5^{\phi^{14}}, 6^{\phi^{24}} )\times F^{(0)}_{\cO_{\rm offset}}(1^{\phi^{12}}, -6^{\phi^{31}}, -5^{\phi^{23}}, -4^{-};q)\,=\,\frac{2\,\la 35\ra \ls 51\rs}{\la 34\ra \la45\ra \ls 54\rs \ls41\rs}\,,
\label{E1}
\\
(ii):&\, i^3\,A(2^{\phi^{23}},3^{\phi^{31}}, 4^{\phi^{14}}, 5^{+}, 6^{\phi^{24}} )\times F^{(0)}_{\cO_{\rm offset}}(1^{\phi^{12}}, -6^{\phi^{31}}, -5^{-}, -4^{\phi^{23}};q)\,=\, \frac{2\, \la46\ra \ls 64\rs}{\la45\ra\la56\ra\ \ls 54\rs \ls65\rs}\,, 
\label{E2}
\\
(iii):& \,i^3\,A(2^{\phi^{23}},3^{\phi^{31}}, 4^{\phi^{14}}, 5^{\phi^{24}}, 6^{+} )\times F^{(0)}_{\cO_{\rm offset}}(1^{\phi^{12}}, -6^{-}, -5^{\phi^{31}}, -4^{\phi^{23}};q)\,=\, \frac{2\,\la 25\ra \ls 51\rs}{\la56\ra\la 62\ra \ls 16\rs \ls65\rs}\,.
\label{E3}
%\end{split}
\end{align}

\noindent 
As explained earlier, the  three  cases corresponding to the opposite helicity assignment of the gluon, which corresponds to  ${F}^{\rm MHV} \times {A}^{\overline{\rm{MHV}}}$ are related to  those discussed above,   ${F}^{\overline{\rm MHV}} \times {A}^{\rm{MHV}}$,  by parity conjugation. The corresponding result is obtained upon   interchanging $\la\cdot ,\cdot \ra \leftrightarrow \ls\cdot ,\cdot \rs$.\\

\begin{figure}[h]
\centering
\includegraphics[width=0.8\linewidth]{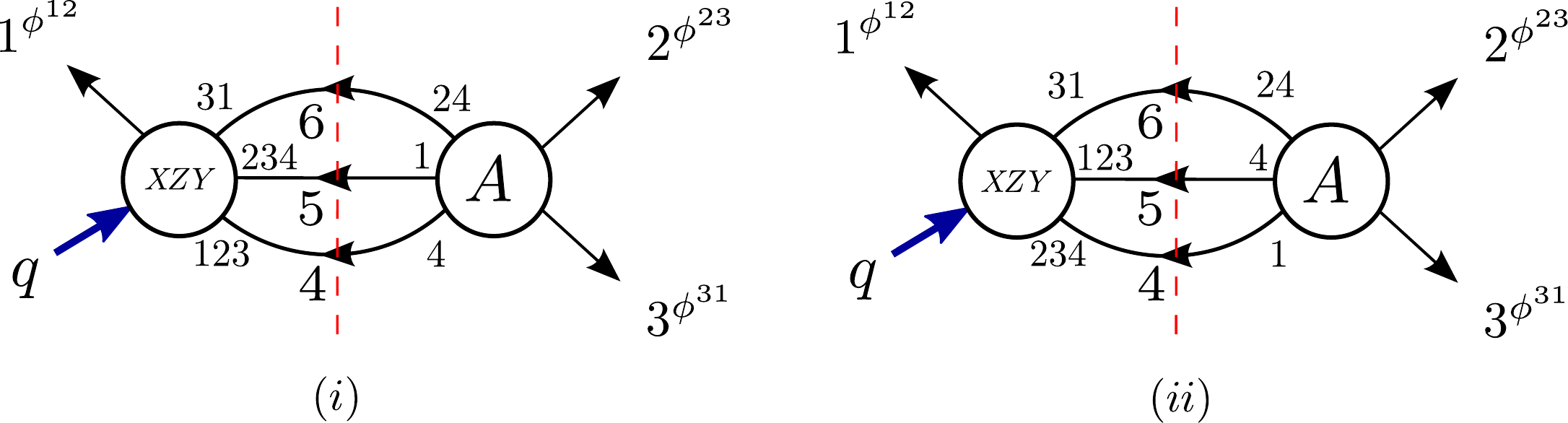}
\caption{\it The first two diagrams with fermions in the loop.  In our conventions, the Yukawa couplings are of  the form, schematically, $\Tr( \phi^{AB} \bar{\psi}_A \bar{\psi}_B)$ and $\Tr( \phi_{AB} {\psi}^A {\psi}^B)$, where $\phi_{AB}$ is related to $\phi^{AB}$ via \eqref{reality}.}
\label{fig:q+1-fermions-first}
\end{figure}

\begin{figure}[h]
\centering
\includegraphics[width=0.8\linewidth]{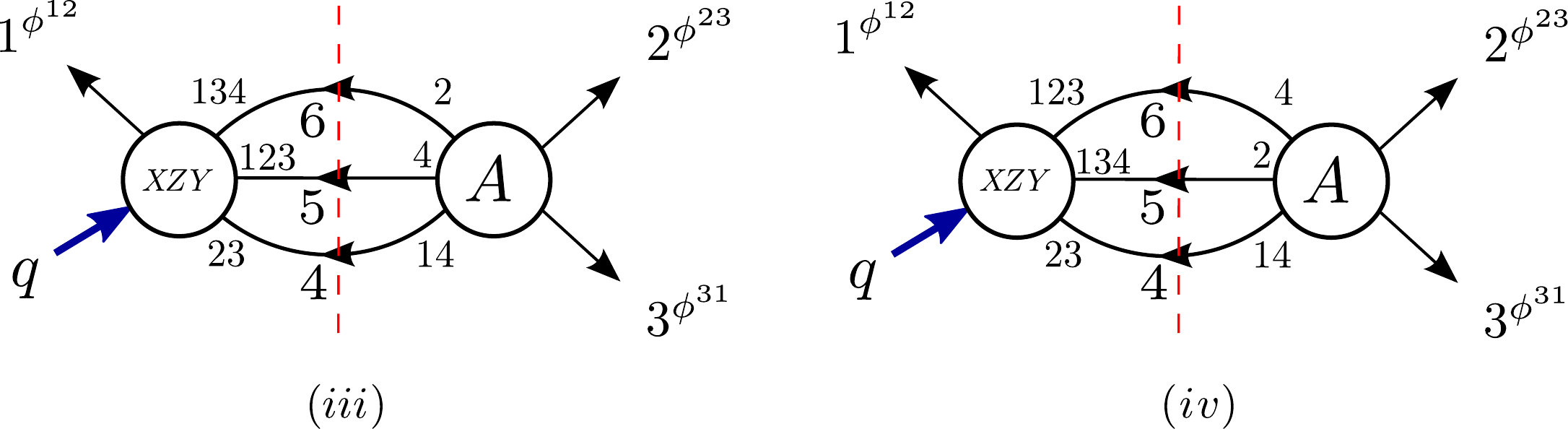}
\caption{\it The remaining two diagrams with fermions in the loop.}
\label{fig:q+1-fermions-second}
\end{figure}

\noindent
{\bf Fermions in the loop.}
Next we consider the situation where two of the loop legs are  fermionic. There are four diagrams corresponding to 
${F}^{\rm \overline{MHV}} \times {A}^{{\rm{MHV}}}$, shown in Figures \ref{fig:q+1-fermions-first} and \ref{fig:q+1-fermions-second}.
%\noindent
The  integrands corresponding to the cuts in Figure \ref{fig:q+1-fermions-first}  are
\begin{align}
(i): \,i^3\,A(2^{\phi^{23}},3^{\phi^{31}}, 4^{\psi^{4}}, 5^{\psi^{1}}, 6^{\phi^{24}} )\times F^{(0)}_{\cO_{\rm offset}}(1^{\phi^{12}}, -6^{\phi^{31}}, -5^{\bar\psi^{234}}, -4^{\bar{\psi}^{123}};q)&\,=\, 2\frac{\la35\ra\la64\ra}{\la34\ra\la56\ra s_{45}}\ ,
\label{EF1}
\\[5pt]
(ii): \,i^3\,A(2^{\phi^{23}},3^{\phi^{31}}, 4^{\psi^{1}}, 5^{\psi^{4}}, 6^{\phi^{24}} )\times F^{(0)}_{\cO_{\rm offset}}(1^{\phi^{12}}, -6^{\phi^{31}}, -5^{\bar{\psi}^{123}}, -4^{\bar\psi^{234}};q)&\,=\, -\frac{2}{s_{45}}\ , 
\label{EF2}
\end{align}

while for the cuts  in Figure~\ref{fig:q+1-fermions-second} we get 
\begin{align}
%\begin{split}
(iii):& \,i^3\,A(2^{\phi^{23}},3^{\phi^{31}}, 4^{\phi^{14}}, 5^{\psi^{4}}, 6^{\psi^{2}} )\times F^{(0)}_{\cO_{\rm offset}}(1^{\phi^{12}}, -6^{\bar\psi^{134}}, -5^{\bar{\psi}^{123}}, -4^{\phi^{23}};q)\,=\, -\frac{2}{s_{56}}\,,
\label{EF4}
\\
(iv):& \,i^3\,A(2^{\phi^{23}},3^{\phi^{31}}, 4^{\phi^{14}}, 5^{\psi^{2}}, 6^{\psi^{4}} )\times F^{(0)}_{\cO_{\rm offset}}(1^{\phi^{12}}, -6^{\bar{\psi}^{123}}, -5^{\bar\psi^{134}}, -4^{\phi^{23}};q)\,=\,  {2\over  s_{56}}\frac{\la25\ra\la46\ra}{\la45\ra\la62\ra}\,.
\label{EF3}
%\end{split}
\end{align}
Again, there are four more diagrams corresponding to ${F}^{\rm MHV} \times {A}^{\overline{\rm{MHV}}}$ which can be obtained using parity conjugation. \\

\noindent
{\bf Combining the terms.}
We can now convert the integrands into traces and dot products and expand them. In doing so, it is useful to notice that the following combination of integrands is particularly simple: 
\beq
\eqref{E1} + \eqref{EF1} + \eqref{EF2} + {1\over 2} \eqref{E2} 
\,=\,\frac{s_{1\ell}}{s_{45}s_{14}} +\frac{s_{13}}{s_{34}s_{14}} -\frac{1}{s_{45}} -\frac{s_{23}
s_{26}}{s_{34}s_{45}s_{56}} -\frac{1}{s_{14}}\ ,
\eeq
where $\ell =-p_4-p_5$ and $p_4$, $p_5$ and $p_6$ are the cut loop momenta.  The corresponding integrals are shown in \eqref{fig:q+1-integrals-first} below. In uplifting the cut expression, we have to pay close attention to the momentum flow: for example, in the expression above $1/s_{14}=1/[2(p_1 \cdot p_4)]$ should be  uplifted to the propagator $-1/(p_1 - p_4)^2$ since  $p_1$ and $p_4$ flow in the same  direction (see Figure \ref{fig:q+1-fermions-first}).  
Keeping these additional signs in mind we arrive at the
the following list of integrals:
\begin{align}\label{fig:q+1-integrals-first}
\hspace{-0.4cm}-s_{1\ell} \times \pic{0.4}{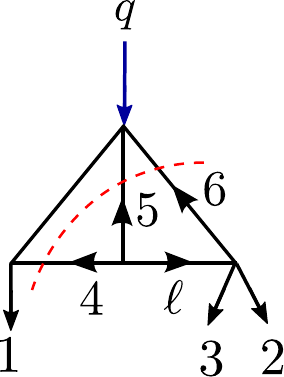} -s_{13}\times \pic{0.4}{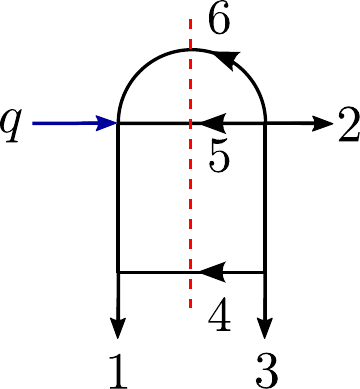}- \pic{0.4}{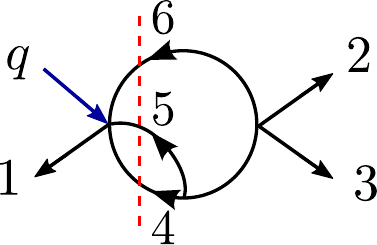}-s_{23}s_{26}\times\pic{0.4}{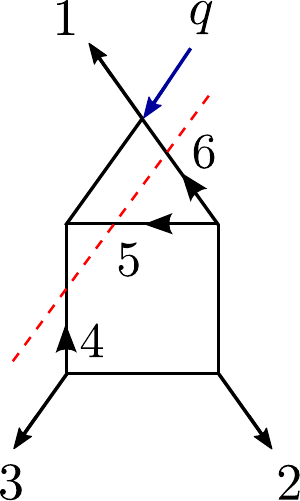}+\quad\pic{0.4}{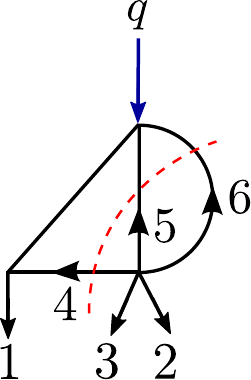}\ ,
\end{align}
\noindent 
Similarly, we single out the following  combination
\beq
\eqref{E3} + \eqref{EF4} + \eqref{EF3} + {1\over 2} \eqref{E2} \ = \ 
\frac{s_{1\ell}}{s_{56}s_{16}}+\frac{s_{12}}{s_{16}s_{26}} -\frac{1}{s_{56}}  -\frac{s_{23}s_{34}}{s_{45}s_{56}s_{26}}  -\frac{1}{s_{16}} \,,
\eeq
where $\ell\,=\,-p_5-p_6$. 
This leads to the integrals shown  below, 
\begin{align}
\label{fig:q+1-integrals-second}
\begin{split}
\hspace{-0.4cm}-s_{1\ell} \times\! \pic{0.4}{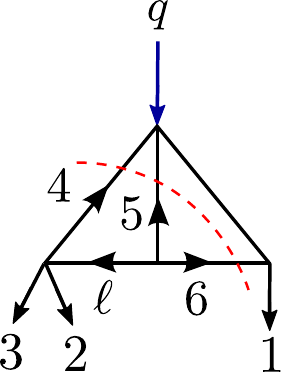}-s_{12}\times\! \pic{0.4}{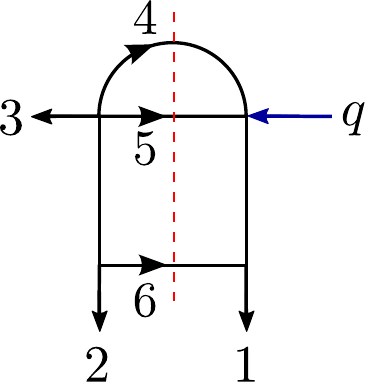}- \pic{0.4}{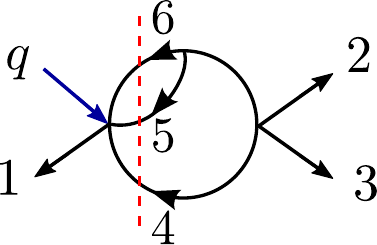}-s_{23}s_{34}\times\pic{0.4}{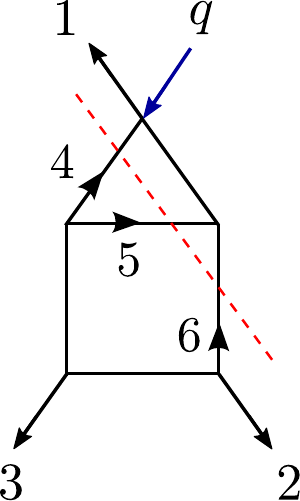}+\quad\pic{0.4}{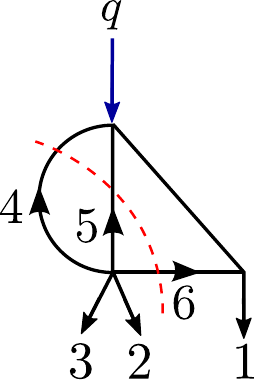}\ .
\end{split}
\end{align} 
The complete  contribution of  the three-particle cut in the $s_{23}$-channel is then obtained by adding \eqref{fig:q+1-integrals-first} and \eqref{fig:q+1-integrals-second}, and multiplying the result by two to take into account the second helicity configuration corresponding to ${F}^{\rm MHV} \times {A}^{\overline{\rm{MHV}}}$.

%%%%%%%%%%%%%%%%

\subsection{Summary and integral reduction}\label{sect:summary-and-reduction}

We now summarise the result of our calculation and  present the result for the form factor of $\cO_B = \Tr ( X[Y, Z])$,  which includes also the half-BPS component $\tilde{\cO}_{\rm BPS} = \Tr ( X\{Y, Z\})$ computed in  \cite{Brandhuber:2014ica} and  quoted in \eqref{BPS}. 
The integral basis  is shown in Table \ref{tab:two-loop basis}.
%\pagebreak
% the [!ht] helps fixing the position of the table
\begin{table}[!ht]
\centering
\begin{tabular}{cccc}
$\pic{0.4}{SOTR}$ & $\pic{0.4}{SOTL}$ & $\pic{0.4}{SOTTBPS}$ & $ \pic{0.4}{Rake2} $ \\[20pt]
$I_1(i)$ & $I_2(i)$ &$I_3(i)$ &$I_4(i)$ \\[10pt]
$\pic{0.4}{Rake3} $ & $\pic{0.4}{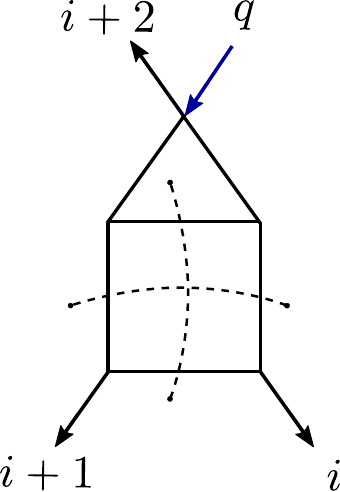}$ & $\pic{0.4}{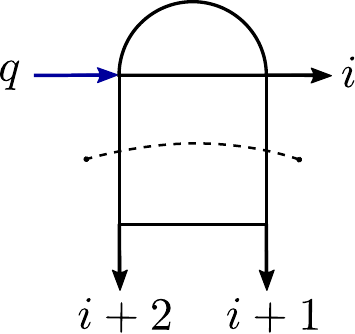}$ & $\pic{0.4}{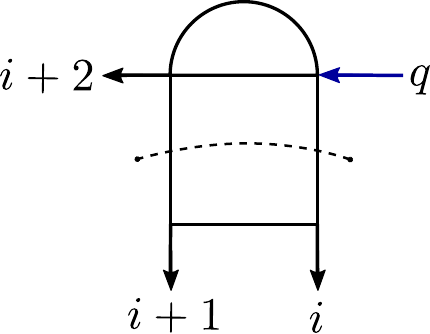}$\\[20pt]
$I_5(i)$ & $I_6(i)$ &$I_7(i)$ &$I_8(i)$ \\[10pt]
$\pic{0.4}{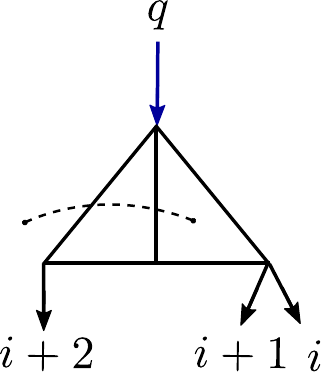} $ & $\pic{0.4}{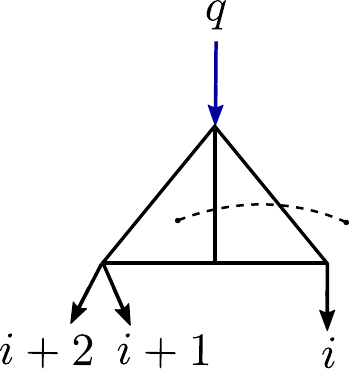} $ & $ \pic{0.4}{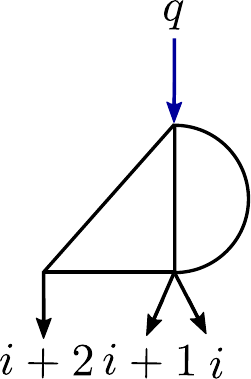} $ & $\pic{0.4}{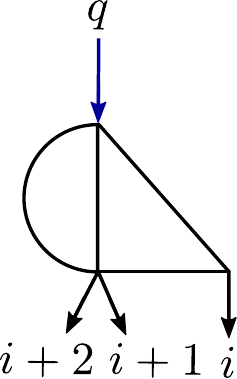} $\\[50pt]
$I_9(i)$ & $I_{10}(i)$ &$I_{11}(i)$ &$I_{12}(i)$ \\[10pt]
$\pic{0.4}{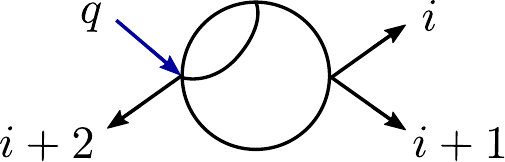}$&$\pic{0.4}{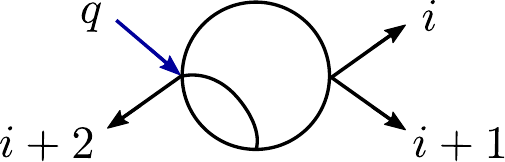}$&$ \pic{0.4}{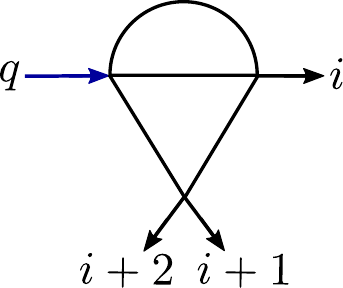}$&$\pic{0.4}{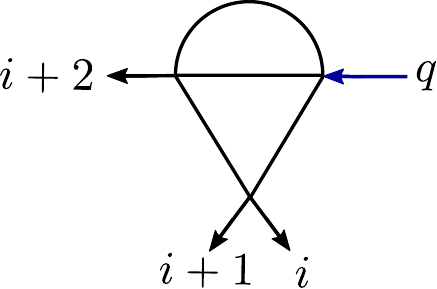}$\\[30pt]
$I_{13}(i)$ & $I_{14}(i)$ &$I_{15}(i)$ &$I_{16}(i)$
\end{tabular}
\caption{\it Integral basis for the two-loop form factor $F_{\cO_B}^{(2)}(1^{\phi^{12}},2^{\phi^{23}},3^{\phi^{31}};q)$. Note that the integrals $\{I_1(i),\ldots,  I_5(i)\}$ correspond precisely to the BPS case, shown in  Eq.~{\rm (3.25)} of {\rm \cite{Brandhuber:2014ica}}. We use the same notation as in {\rm \cite{Brandhuber:2014ica}}: factors of $s_{ij}$/$s_{ijk}$ in the numerators are denoted by a dashed line intersecting two/three lines whose sum of momenta square to the corresponding kinematic invariant.}
\label{tab:two-loop basis}
\end{table}
In terms of this basis, the two-loop minimal form factor of $\cO_B$ is given by
\begin{align}
\begin{split}
\label{final-result}
F_{\cO_B}^{(2)}(1^{\phi^{12}},2^{\phi^{23}},3^{\phi^{31}};q)=&-\sum_{i=1}^4 I_i(1) \ +I_5(1)  
\ -\  2\times\Big[ \sum_{i=6}^{10}   I_i(1) \ -I_{11}(1)\ -I_{12}(1)\\
&+I_{13}(1)\ +I_{14}(1)\Big]\ -4\times\Big[I_{15}(1)+I_{16}(1)\Big]\ \\
& + \ \text{cyclic}(1,2,3)\ .
\end{split}
\end{align}
Some of the integrals appearing in \eqref{final-result} are  master integrals and we can proceed to substitute their expressions from \cite{Gehrmann:1999as,Gehrmann:2000zt}.  The remaining ones will be reduced using a particular integration-by-parts algorithm implemented in the {\texttt{Mathematica}} package {\texttt{LiteRed}} \cite{Lee:2012cn,Lee:2013mka}. Using this package we find the following reductions: 
\begin{align}
\begin{split}
&\pic{0.45}{SOTT}\,=\,\frac{4(\epsilon-1)(3\epsilon-2)(3\epsilon-1)}{\epsilon^2 s_{i\,i+1}(2\epsilon-1)}\pic{0.45}{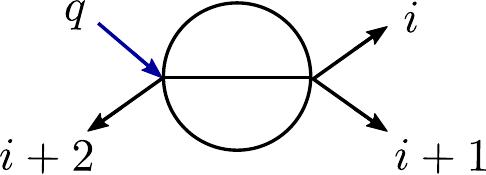}\\ 
&-\frac{2(3\epsilon-1)}{\eps}\pic{0.45}{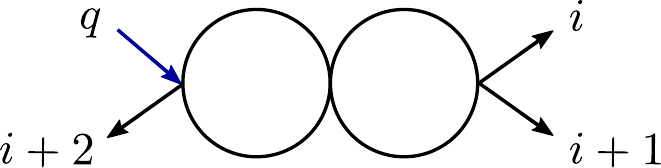}-\frac{2(\epsilon-1)}{\eps}\pic{0.45}{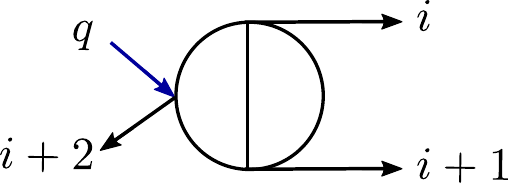}\ ,
\end{split}
\end{align}
\begin{align}
\begin{split}
&\pic{0.45}{LMRake}\,=\,\frac{(3\eps-2)[s_{i\,i+1}\eps+(2\eps-1)(s_{i\,i+2}+s_{i+1\,i+2})]}{\eps^2(s_{i\,i+2}+s_{i+1\,i+2})s_{i\,i+1}}\pic{0.45}{sunset2}\\
&-\frac{2\eps-1}{\eps}\;\pic{0.45}{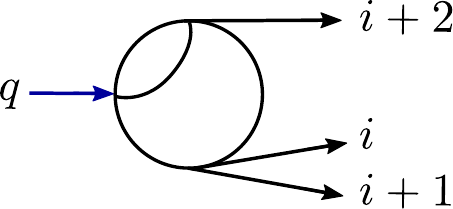}-\frac{3\eps-2}{\eps(s_{i\,i+2}+s_{i+1\,i+2})}\;\pic{0.45}{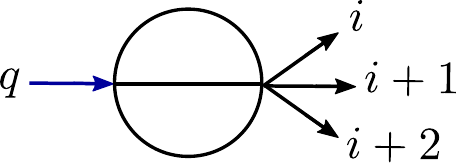}\ ,
\end{split}
\end{align}
\begin{align}
&\pic{0.45}{RA}\,=\,\frac{3\eps-2}{2\eps(s_{i\,i+2}+s_{i+1\,i+2})}\Bigg(\;\pic{0.45}{sunset2}-\;\pic{0.45}{sunset3}\;\Bigg)\,,\\[15pt]
&\pic{0.45}{teardrop-up}\,=\,\frac{3\eps-2}{2\eps\, s_{i\,i+1}}\; \pic{0.45}{sunset2}\ .
\end{align}
\\
These reduced integrals, with expressions known from \cite{Gehrmann:1999as,Gehrmann:2000zt}, can then be plugged into \eqref{final-result} to give the final result of the two-loop form factor $F_{\cO_B}^{(2)}(1^{\phi^{12}},2^{\phi^{23}},3^{\phi^{31}};q)$. We refrain from writing the full expression for this form factor at present due to its considerable length. Instead, we  consider next a much simpler quantity obtained from a standard subtraction of the IR singularities~--~the remainder function.

\section{Two-loop remainder    of $\langle\bar{X}\bar{Y}\bar{Z} |\,  \Tr X[Y, Z] |0\rangle$}\label{sect:TwoLoopRemainder}
\subsection{Definition of the  remainder }

Two-loop remainder functions for  the form factor of a generic operator $\cO$   were introduced in \cite{Brandhuber:2012vm} similarly to the amplitude remainder function \cite{0309040, Bern:2005iz}, 
\beq
\label{eq:remainder}
\cR^{(2)}_{\O} \ := \    F_{\O}^{(2)}(\epsilon )\, - \, {1\over 2} \big( F_{\O}^{(1)} (\epsilon) \big)^2 -  f^{(2)} (\epsilon)\  F_{\O}^{(1)} ( 2 \epsilon ) - C^{(2)}
\, + \cO (\epsilon ) \
\ , 
\eeq
where 
$f^{(2)} (\eps):= -2(  \zeta_2 + \eps \, \zeta_3 + \eps^2 \, \zeta_4)$ and
$C^{(2)}=4 \zeta_4$. 
As in \cite{Brandhuber:2012vm,Brandhuber:2014ica}, the function $f^{(2)} (\epsilon) $ is the same as for amplitudes \cite{0309040, Bern:2005iz}. Note that we have defined the remainder by taking out a power of
\beq
  {g^2 N  e^{- \epsilon \gamma_{\rm E}}\over  (4 \pi )^{2-\epsilon }} = a (4 \pi e^{- \gamma_{\rm E}})^{\epsilon}
  \ 
 \eeq
per loop, where $a$ is our 't Hooft coupling, defined in \eqref{Hooft}. We also observe  that in general
we would define the remainder for the helicity-blind ratio $F_{\O}^{(2)}/F_{\O}^{(0)}$ as in \cite{Brandhuber:2012vm} but in this particular case this is not necessary since the tree-level form factor is equal to one. An important aspect of this procedure \cite{Brandhuber:2012vm,0309040,Bern:2005iz} is that it removes the universal IR divergences of the result. In the case of protected operators this gives a finite remainder while in the present case, where we consider a bare, unprotected operator, we are still left with UV divergences. In Section \ref{mixing} we will determine the appropriate renormalised operators and form factors that have  a UV and IR finite remainder function. Here however we wish to take a first look at the IR-finite, but UV-divergent remainder function of the form factor $\langle\bar{X}\bar{Y}\bar{Z} |\,  \Tr (X[Y, Z]) |0\rangle$.

Using the decomposition \eqref{eq:-2XZY}, 
the remainder function splits into a term formed completely by the form factor of $\tilde{\O}_{\rm BPS}$ and a piece which contains  mixed terms involving  $\tilde{\O}_{\rm BPS}$ and $\O_{\rm offset}$, which we denote by $\cR^{(2)}_{{\text{non-BPS}}}$:
\beq\label{split-remainder}
\cR_{\O_B}^{(2)} \ = \ \cR^{(2)}_{\rm BPS}  \, + \,  \cR^{(2)}_{{\text{non-BPS}}}\ , 
\eeq
where
\begin{align}
\cR^{(2)}_{\rm BPS}\  &= \  F^{(2)}_{\tilde{\O}_{\rm BPS}}(\eps)  \, - \, {1\over 2} \big( F^{(1)}_{\tilde{\O}_{\rm BPS}}  (\epsilon) \big)^2 -  f^{(2)} (\epsilon)\  F^{(1)}_{\tilde{\O}_{\rm BPS}}  ( 2 \epsilon ) - C^{(2)}\ ,\\
\label{non-BPS-remainder-def}
\cR^{(2)}_{{\text{non-BPS}}}\  &= \  F^{(2)}_{\cO_{\rm offset}} (\eps) \, - \, F^{(1)}_{\cO_{\rm offset}}\Big( {1\over 2} F^{(1)}_{\cO_{\rm offset}}  +  F^{(1)}_{\tilde{\O}_{\rm BPS}}\Big) (\epsilon) -   f^{(2)} (\epsilon)\  F^{(1)}_{\cO_{\rm offset}} ( 2 \epsilon ) \ .
\end{align}
The remainder of the half-BPS operator $\Tr ( X^3)$ was computed in  Eq.~(4.21) of \cite{Brandhuber:2014ica} and is identical to the BPS remainder appearing here. It is given by a function of uniform transcendentality equal to four, written in terms of classical polylogarithms only. Explicitly, its expression is
\begin{align}
\label{eq:remainderBPS}
\begin{split}
\cR^{(2)}_{\rm BPS} \ := \ & 
 -\frac{3}{2}\, \text{Li}_4(u)+\frac{3}{4}\,\text{Li}_4\left(-\frac{u v}{w}\right)  
-\frac{3}{2}\log(w) \, \text{Li}_3 \left(-\frac{u}{v} \right)   
+ \frac{ 1}{16}  {\log}^2(u)\log^2(v)
\\
&
+{\log^2 (u) \over 32} \Big[ \log^2 (u) - 4 \log(v) \log(w) \Big]  
+ {\zeta_2 \over 8 }\log(u) [  5\log(u)- 2\log (v) ]
 \\
&
+ {\zeta_3 \over 2} \log(u) +\frac{7}{16}\, \zeta_4  + {\rm perms}\, (u,v,w) \ , 
\end{split}
\end{align}
where
\begin{align}
\label{eq:uvw}
u\, =\, \frac{s_{12}}{q^2}, \qquad v\, =\,\frac{s_{23}}{q^2},\qquad w\, =\, \frac{s_{31}}{q^2},\qquad\qquad u+v+w=1\ .
\end{align}
The new part is the non-BPS remainder defined in \eqref{non-BPS-remainder-def}. It is IR finite, but it still has UV divergences due to the fact that the operator inserted is not protected. 
Interestingly, it is given by a sum of functions of  transcendentality ranging from three to zero, with no term with maximal transcendentality: 
\beq
\label{eq:remainder-UV-coefficient}
\cR^{(2)}_{\text{non-BPS}}\ \ = \ 
{c\over \epsilon}  \ + \ 
\sum_{i=0}^3\cR^{(2)}_{{\text{non-BPS}};3-i}\  \ ,  
\eeq
where the subscript $m$ in $\cR^{(2)}_{\text{non-BPS};m}$ denotes the degree of transcendentality of the corresponding  term. For the coefficient of the  UV pole we find 
\beq
\label{pisquared}
c \  = \ 18 - \pi ^2
\ . 
\eeq
The expression arising from replacing the integral functions appearing in the two-loop form factor with the explicit results of 
\cite{Gehrmann:1999as,Gehrmann:2000zt}
can be considerably simplified using the concept of the symbol  of a transcendental function \cite{Goncharov:2010jf}, while  beyond-the-symbol terms can be fixed numerically and/or analytically. At transcendentality three, we are guaranteed that the whole result can be written in terms of classical polylogarithms only, and hence this procedure is very simple to carry out. We find that the symbol of 
$\cR^{(2)}_{{\text{non-BPS}};3}$ is 
\beq
\label{eq:symbol3}
\mathcal{S}_{3}^{(2)}(u,v,w)\, =\, 
- 2\Big[  u\,\otimes (1-u)\,\otimes\, {u\over 1-u}
\, +u\, \otimes \,u\otimes\, {v\over 1-u}
\, + \, u\,\otimes\, v\otimes\, {uv\over w^2} 
\Big]  
\,   + \,\, \text{perms}\, (u,v,w)\ ,
\eeq
while for the integrated expression (including beyond-the-symbol terms) we get
\beqa
\cR^{(2)}_{{\text{non-BPS}};3}&=& 
2 \Big[ 
\text{Li}_3(u)
+ \text{Li}_3(1-u)
\Big]
- {1\over 2} \log^2(u) \log {v w\over (1-u)^2} 
+{2\over 3} \log (u) \log (v) \log (w)
\nonumber \\
&+& 
 {2\over3} \,  \zeta_3 \, + \, 2\,  \zeta_2 \log ( - q^2) \, + \, 
\text{perms}\, (u,v,w)\ .
  \eeqa
The transcendentality-two part of the remainder can also  be simplified slightly. 
A short calculation leads to the expression 
\beq
\label{rem-transc-2}
\cR^{(2)}_{{\text{non-BPS}};2} \ = \ -12 \Big[ \text{Li}_2(1-u)+\text{Li}_2(1-v)+ \text{Li}_2(1-w)\Big]
-2 \log ^2(u v
   w)\, +\, 36 \, \zeta_2\ . 
   \eeq
Finally, for the transcendentality-one and zero terms we have
\beqa
\cR^{(2)}_{{\text{non-BPS}};1} & = & -12\,  \log (uvw)   -36\,  \log (-q^2 ) 
\, , 
\\
\label{rem-transc-0}
\cR^{(2)}_{{\text{non-BPS}};0}& = & 126
\ . 
\eeqa
Before concluding this section we would like to make two observations on the results we have derived here.

\begin{itemize}
\item[{\bf 1.}]
First, we observe  that the $- \pi^2$ term in \eqref{pisquared} comes from the last term on the right-hand side of \eqref{non-BPS-remainder-def}. It amounts to introducing a spurious UV divergence in the remainder arising from the bubbles contained in the term $F^{(1)}_{\cO_{\rm offset}} ( 2 \epsilon )$. For the sake of extracting the correct UV divergences and studying the mixing, this term must be omitted, see   Section \ref{mixing} for this discussion.

\item[{\bf 2.}]
We stress the usefulness of  the decomposition \eqref{eq:-2XZY} and \eqref{split-remainder}, which has the great advantage of separating out completely the terms of maximal transcendentality from the rest. 
This is in line with the findings of  \cite{Loebbert:2015ova}, where it was observed in the $SU(2)$ sector that a 
the finite remainder densities introduced there, and corresponding to different ``shuffling" for the $R$-symmetry fields flavours, have a highest degree of transcendentality equal to $4-s$ with $s$ being the shuffling in that remainder density. In the present case, in a different sector, the operator $\Tr (XZY)$ is associated with an external state $\langle \bar{X} \bar{Y} \bar{Z} |$, which corresponds to $s\!=\!1$. Indeed we find that the corresponding remainder is composed of terms with transcendentality ranging from three to zero. 

\end{itemize}

\subsection{A connection to the remainder densities in the $SU(2)$ sector} 
We now establish a connection between the (UV-finite part of the) non-BPS remainder $\cR^{(2)}_{\text{non-BPS}}$
and the remainder densities which have appeared in  \cite{Loebbert:2015ova} in connection with the calculation of the dilatation operator in the $SU(2)$ sector. This  is a closed subsector of $SU(2|3)$ and operators are built out of the complex scalars $X$ and $Y$ defined earlier in \eqref{eq:XYZ-phi}.
Two observations are in order here. Firstly we note that the remainder densities studied in  \cite{Loebbert:2015ova} correspond to operators  which are products of fields without the trace. 
Secondly, the operator we are considering is part of the larger $SU(2|3)$ sector, hence we should not expect to find similarities with results obtained in smaller sectors. In particular, in the $SU(2|3)$ sector the spin chain becomes {\it dynamic} i.e. the  number of spin sites can  fluctuate due to length-changing interactions, something which cannot occur in the $SU(2)$ sector. We will see in Section \ref{mixing} that this is important for the renormalisation of the form factor of $\cO_B$. 

It was found in   \cite{Loebbert:2015ova}  that there are only three independent finite remainder densities, denoted in that paper as 
$\big(R_i^{(2)}\big)^{XXX}_{XXX}$, 
$\big(R_i^{(2)}\big)^{XYX}_{XXY}$,  and   
$\big(R_i^{(2)}\big)^{YXX}_{XXY}$. 
The first density,  $\big(R_i^{(2)}\big)^{XXX}_{XXX}$, has uniform transcendentality equal to four and is identical to the half-BPS remainder computed in \cite{Brandhuber:2014ica}. $\big(R_i^{(2)}\big)^{XYX}_{XXY}$ contains terms of transcendentality ranging from three to zero, while $\big(R_i^{(2)}\big)^{YXX}_{XXY}$ contains terms of transcendentality two, one and zero. The index $i$ denotes the spin chain site, and the remainder densities depend on the  three variables 
\beq
u_i \ = \ {s_{i \, i +1} \over s_{i \, i +1 \, i+2}}\,  , \qquad 
v_i \ = \ {s_{i +1\,  i+2} \over s_{i \, i +1 \, i+2}}\,   ,\qquad 
w_i \ = \ {s_{i \, i +2} \over s_{i \, i +1 \, i+2}}\ ,   
\eeq
as well as on  $s_{i \, i +1}$, $s_{i+2 \, i +2}$, $s_{i \, i +2}$ and $s_{i \, i +1 \, i+2}$ separately. 

We have observed an  interesting connection between these remainder densities  and our non-BPS remainder, namely
\beqa
{1\over 2} \cR^{(2)}_{{\text{non-BPS}};3}  &=&  -\sum_{S_3} \left.\big(R_i^{(2)}\big)^{XYX}_{XXY}\right|_3 \  + \ 6 \, \zeta_3\ , 
\nonumber \\
{1\over 2} \cR^{(2)}_{{\text{non-BPS}};2} &=&
-\sum_{S_3}\left.  \Big[ \big(R_i^{(2)}\big)^{XYX}_{XXY} - \big(R_i^{(2)}\big)^{YXX}_{XXY} \Big]  \right|_2 \ +\  5 \pi^2\ , 
\nonumber \\
{1\over 2} \cR^{(2)}_{{\text{non-BPS}};1} &=&
-\sum_{S_3}\left.  \Big[ \big(R_i^{(2)}\big)^{XYX}_{XXY} - \big(R_i^{(2)}\big)^{YXX}_{XXY} \Big]  \right|_1 \ , 
\nonumber \\
{1\over 2} \cR^{(2)}_{{\text{non-BPS}};0}&=&
-\sum_{S_3}\left.  \Big[ \big(R_i^{(2)}\big)^{XYX}_{XXY} - \big(R_i^{(2)}\big)^{YXX}_{XXY} \Big]  \right|_0 \ , 
\eeqa
where $\left.f\right|_m$ denotes the transcendentality-$m$ part of the function $f$, the remainder densities are evaluated  with the replacements $(u_i, v_i, w_i) \to (u, v, w)$, and $S_3$ denotes  permutations of $(u, v, w)$.  It would be very interesting to explain this almost perfect coincidence of these a priori unrelated quantities.

%%%%%%%%%%%

\section{One-loop non-minimal form factor  $\langle \bar X\bar Y\bar Z | \frac{1}{2}\Tr \psi^\alpha \psi_\alpha | 0 \rangle$}
\label{sect:NonMinimal}

In this section we compute one of the off-diagonal entries of the matrix of form factors \eqref{eq:FF-matrix}, namely $F^{(1)}_{\cO_F}(1^{\phi^{12}},2^{\phi^{23}},3^{\phi^{31}};q)$, where  $\cO_F =(1/2) \Tr (\psi^\alpha \psi_\alpha)$.
Note that $\cO_F$ is defined in a way that its minimal tree-level form factor $ \la \bar\psi^{123}(1) \bar\psi^{123}(2)  | \cO_F(0) | \, 0\,  \ra$
 is equal to $\langle 21\rangle$.

In order to do so we construct the one-loop integrand by considering two-particle  cuts in the  $q^2$ and $s_{23}$ channels. We will find  that the result is IR finite as it should be since this form factor does not exist at tree level.  However, UV divergences are expected reflecting the mixing between $\cO_B$ and $\cO_F$.   This will be studied in detail in Section \ref{mixing}.
\subsection{Two-particle cut in the $q^2$-channel}
\begin{figure}[h]
\centering
\includegraphics[width=0.5\linewidth]{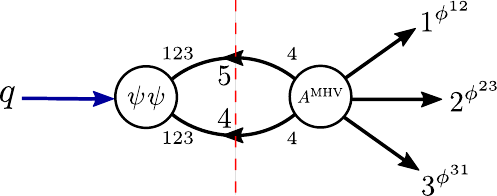}
\caption{\it Two-particle cut of the non-minimal form factor $F^{(1)}_{\cO_F}$ with external state $\langle \bar{X}\bar{Y}\bar{Z} |$. }
\label{fig:q-cut}
\end{figure}

\begin{figure}[h]
\centering
\includegraphics[width=0.25\linewidth]{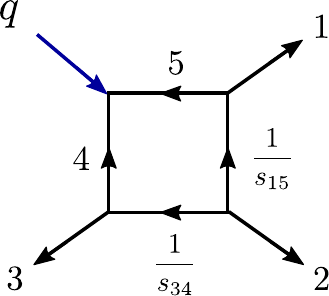}
\caption{\it The integral topology that appears in the $q^2$-channel two-particle cut. For future convenience we indicate explicitly the uncut propagators.}
\label{fig:q-cut-result}
\end{figure}

\noindent
We start by computing the $q^2$-channel of the form factor $F_{\cO_F}^{(1)}(1^{\phi^{12}},2^{\phi^{23}},3^{\phi^{31}};q)$. This is shown in Figure \ref{fig:q-cut} and is given by
\begin{align}
\label{eq:q^2-cut}
\begin{split}
&F_{\cO_F}^{(1)}(1^{\phi^{12}},2^{\phi^{23}},3^{\phi^{31}};q)\Big|_{2,q^2}\,= i^2 F^{(0)}_{\cO_F}(-5^{\bar{\psi}^{123}},-4^{\bar{\psi}^{123}};q) \times A(1^{\phi^{12}},2^{\phi^{23}},3^{\phi^{31}},4^{\psi^4},5^{\psi^4})\,\\
&=\,-i \b{45} \times \frac{\b{13}}{\b{34}\b{51}}
\,=\,-\frac{i}{2} \Big(\frac{s_{34}s_{15}+s_{45}s_{13}-s_{14}s_{35}}{s_{34}s_{15}}\Big)\ .
\end{split}
\end{align}
The corresponding topology is the   box shown in Figure \ref{fig:q-cut-result}.

\noindent
We now rewrite the numerators in  \eqref{eq:q^2-cut}  using 
\begin{align}
s_{45}\,=\,s_{123}\,,\quad s_{14}\,=\,-(s_{12}+s_{13}+s_{15})\,,\quad s_{35}\,=\,-(s_{31}+s_{32}+s_{34})\ , 
\end{align}
which follow from momentum conservation $\sum_{i=1}^5 p_i\!=\!0$ and the cut conditions $p_4^2~=~p_5^2~=~0$.
Doing so \eqref{eq:q^2-cut} becomes
\begin{align}
\label{eq:q^2-cut-result}
\begin{split}
&F_{\cO_F}^{(1)}(1^{\phi^{12}},2^{\phi^{23}},3^{\phi^{31}};q)\Big|_{2,q^2}\,=\,\frac{i}{2}\Big(\frac{s_{12}s_{23}}{s_{34}s_{15}}+\frac{s_{13}+s_{23}}{s_{34}}+\frac{s_{12}+s_{13}}{s_{15}}\Big) \\
\,&=\,\frac{i}{2}\Big[s_{12}s_{23}\times\pic{0.4}{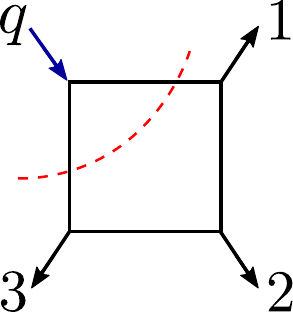}+(s_{13}+s_{23})\times\pic{0.4}{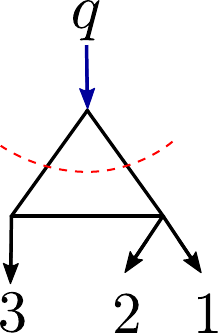}+(s_{12}+s_{13})\times\pic{0.4}{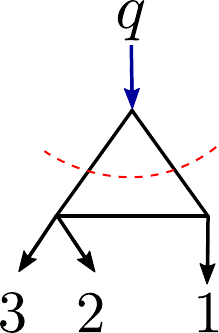}\Big]\ .
\end{split}
\end{align}
Note that in this cut no UV-divergent integrals have appeared and we have to add two additional contributions from cyclic permutations of the external particles.

\subsection{Two-particle cut in the $s_{23}$-channel}

\begin{figure}[h]
\centering
\includegraphics[width=\linewidth]{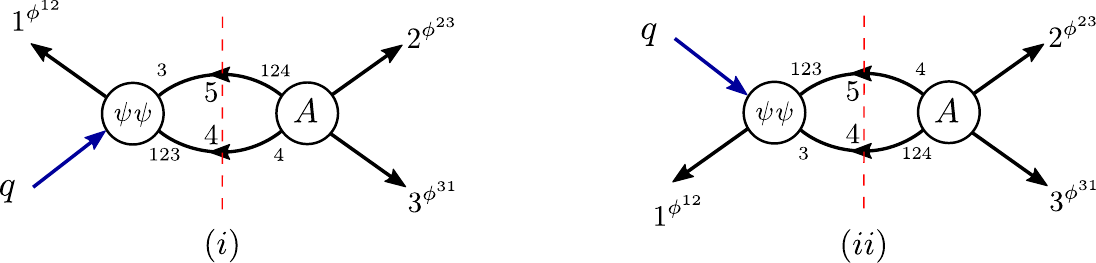}
\caption{\it Two diagrams entering the two-particle cut in the $s_{23}$-channel.}
\label{fig:q-p1-cut}
\end{figure}

We now compute the two-particle cut of $F^{(1)}_{\cO_F}(1^{\phi^{12}},2^{\phi^{23}},3^{\phi^{31}};q)$ in  the $s_{23}$-channel. 
There are two possible diagrams to consider, shown in Figure \ref{fig:q-p1-cut}. 

\begin{figure}[h]
\centering
\includegraphics[width=0.7\linewidth]{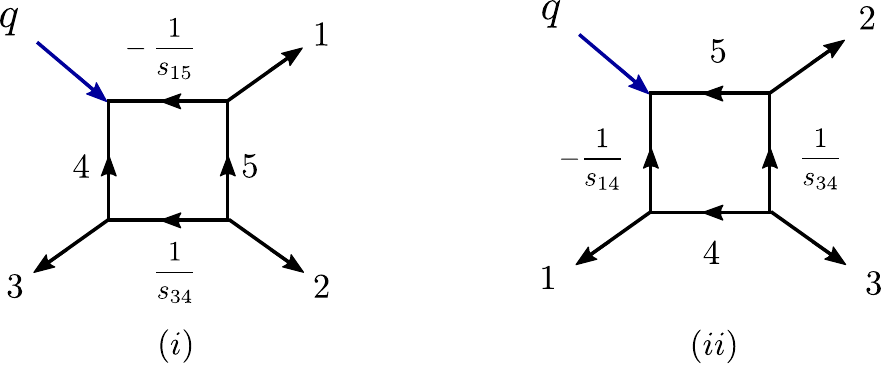}
\caption{\it Master topologies generated by the two diagrams of Figure \ref{fig:q-p1-cut}, respectively. The uncut propagators are explicitly shown in order to bookkeep their sign reflecting the momentum flow. For the coefficient of the box integral, only the diagram on the left can be compared with the box detected in the $q^2$-cut of Figure \ref{fig:q-cut} due to the ordering of external legs.}
\label{fig:master-topologies}
\end{figure}

\noindent 
These two diagrams give rise to   the master topologies shown in Figure \ref{fig:master-topologies}, with corresponding numerators determined by the cuts. 

\begin{figure}[h]
\centering
\includegraphics[width=0.5\linewidth]{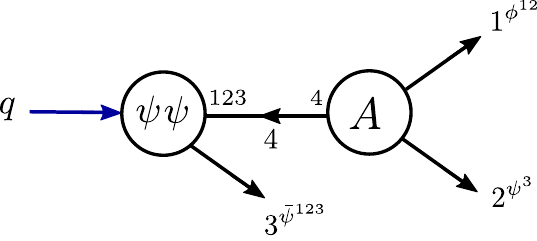}
\caption{\it A factorisation diagram of the non-minimal form factor  $F^{(0)}_{\cO_F}(1^{\phi^{12}},2^{\psi^3},3^{\bar{\psi}^{123}};q)$ featuring in the two-particle cut of $F^{(1)}_{\cO_F}(1^{\phi^{12}},2^{\phi^{23}},3^{\phi^{31}};q)$ in the $s_{23}$-channel.}
\label{fig:non-minimal}
\end{figure}

\noindent In the cuts  we need the tree-level non-minimal form factors $F^{(0)}_{\cO_F}(1^{\phi^{12}},2^{\psi^3},3^{\bar{\psi}^{123}};q)$ and $F^{(0)}_{\cO_F}(1^{\phi^{12}},2^{\psi^{123}},3^{\bar{\psi}^{3}};q)$. The first of them has only one possible factorisation diagram corresponding to a fermion splitting into an anti-fermion and a scalar, as shown in Figure \ref{fig:non-minimal}.

\noindent From this factorisation diagram we can infer the expression for the tree-level form factor, which is given by
\begin{align}
F^{(0)}_{\cO_F}(1^{\phi^{12}},2^{\psi^3},3^{\bar{\psi}^{123}};q)\,=\, F^{(0)}_{\cO_F}(-4^{\bar{\psi}^{123}},3^{\bar{\psi}^{123}};q)\times \frac{i}{s_{12}}\times A^{\MHVb}(1^{\phi^{12}},2^{\psi^3},4^{\psi^4})\ . 
\end{align}
The anti-MHV amplitude can be easily determined using parity,
\begin{equation}
A^{\MHVb}(1^{\phi^{12}},2^{\psi^3},4^{\psi^4})\,=\, - \big[A^{\rm MHV}(1^{\phi^{34}},2^{\bar\psi^{124}},4^{\bar{\psi}^{123}})\big]^*\,=\,i\,[24]\ .
\end{equation}
Using $p_4=-(p_1+p_2)$ we obtain the result 
\begin{align}
\label{eq:non-minimal-ff}
F^{(0)}_{\cO_F}(1^{\phi^{12}},2^{\psi^3},3^{\bar{\psi}^{123}};q)\,=\, \frac{[21]\b{13}}{s_{12}} \ .
\end{align}

\noindent We now compute the two diagrams of Figure \ref{fig:q-p1-cut} separately.

\subsubsection*{Diagram $\boldsymbol{(i)}$}
This diagram  is given by
\begin{align}
\label{eq:cut-a}
\begin{split}
F_{\cO_F}^{(1)}(1^{\phi^{12}},2^{\phi^{23}},3^{\phi^{31}};q)\Big|_{2,s_{23},(i)}\,=\,
&-i^2\, F^{(0)}_{\cO_F}(1^{\phi^{12}},-5^{\psi^3},-4^{\bar{\psi}^{123}};q)\times A^{\rm MHV}(2^{\phi^{23}},3^{\phi^{31}},4^{\psi^4},5^{\bar\psi^{124}})
\\=\,
&-\frac{i}{2}\Big(\frac{s_{14}s_{35}+s_{34}s_{15}-s_{13}s_{45}}{s_{15}s_{34}}\Big)\ .
\end{split}
\end{align}

\noindent Using $p_2+p_3+p_4+p_5=0$ and $p_4^2=p_5^2=0$ on the cut, we can substitute
\begin{equation}
\label{eq:replacements-q-p1}
s_{45}\,=\,s_{23}\,,\quad s_{35}\,=\,-(s_{34}+s_{32})\,,\quad s_{14}\,=\,-(s_{12}+s_{13}+s_{15})\ ,
\end{equation} 
thus \eqref{eq:cut-a} becomes
\begin{align}
\begin{split}
\label{eq:cut2a}
&F_{\cO_F}^{(1)}(1^{\phi^{12}},2^{\phi^{23}},3^{\phi^{31}};q)\Big|_{2,s_{23},(i)}\,=\,-\frac{i}{2}\left[2 + \frac{s_{12}+s_{13}}{s_{15}} + \frac{s_{23}}{s_{34}} + \frac{s_{12}s_{23}}{s_{15}s_{34}} \right]\\
=\, & \frac{i}{2} \Big[ -2\times\pic{0.4}{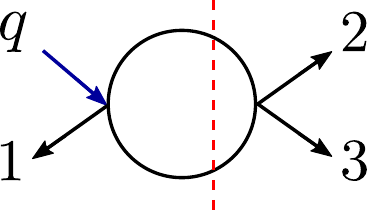} +(s_{12}+s_{13})\times\pic{0.4}{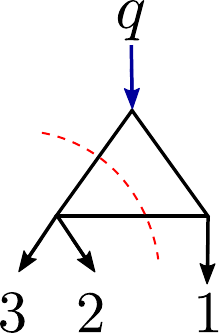} - s_{23} \times \pic{0.4}{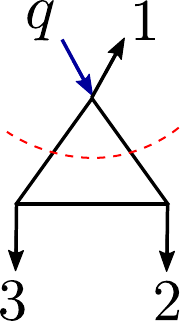}+s_{12}s_{23}\times\pic{0.4}{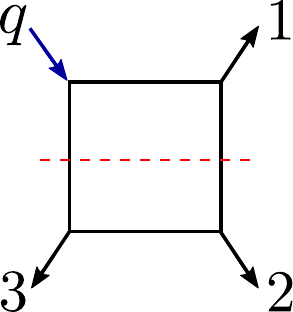}\Big]\ .
\end{split}
\end{align}
Note that when the cut-integrals are uplifted to full Feynman integrals $1/s_{15}$,  has to be replaced by $-1/(p_1-p_5)^2$ due to the momentum flow, according to Figure \ref{fig:master-topologies}$(i)$.

\subsubsection*{Diagram $\boldsymbol{(ii)}$}
For diagram $(ii)$ we need the form factor
\begin{align}
\label{eq:ff2b}
\begin{split}
&F^{(0)}_{\cO_F}(1^{\phi^{12}},-5^{\bar{\psi}^{123}},-4^{\psi^3};q)\,=\,F^{(0)}_{\cO_F}(1^{\phi^{12}},-4^{\psi^3},-5^{\bar{\psi}^{123}};q)\,=\,\frac{[41]\b{15}}{s_{14}}\ .
\end{split}
\end{align}
Its expression is given by 
\begin{align}
\label{eq:diagb}
\begin{split}
&F_{\cO_F}^{(1)}(1^{\phi^{12}},2^{\phi^{23}},3^{\phi^{31}};q)\Big|_{2,s_{23},(ii)}\,=\,-i^2 F^{(0)}_{\cO_F}(1^{\phi^{12}},-5^{\bar{\psi}^{123}},-4^{\psi^3};q)\times A^{\rm MHV}(2^{\phi^{23}},3^{\phi^{31}},4^{\bar\psi^{124}},5^{\psi^{4}})\\[5pt]
\,=\,&i \frac{[41]\b{15}}{s_{14}}\times \left(\frac{\b{24}}{\b{52}}\right)
%\,=\,-i \frac{[14]\b{15}\b{24}[25]}{s_{14}s_{25}} 
=\, -i \frac{\Tr_-(1524)}{s_{14}s_{25}}\,=\, i \frac{\Tr_-(1534)}{s_{14}s_{34}}\ ,
\end{split}
\end{align}
where we used momentum conservation in the last step. Expanding the trace and using a set of replacements similar to \eqref{eq:replacements-q-p1},
\begin{equation}
\label{eq:replacements-q-p2}
s_{45}\,=\,s_{23}\,,\quad s_{35}\,=\,-(s_{34}+s_{32})\,,\quad s_{15}\,=\,-(s_{12}+s_{13}+s_{14})\ ,
\end{equation}
we arrive at the result
\begin{align}
\label{eq:cut2b}
\begin{split}
&F_{\cO_F}^{(1)}(1^{\phi^{12}},2^{\phi^{23}},3^{\phi^{31}};q)\Big|_{2,s_{23},(ii)}\,=\, -\frac{i}{2}\left[2 + \frac{s_{12}+s_{13}}{s_{14}}+\frac{s_{23}}{s_{34}}+\frac{s_{13}s_{23}}{s_{14}s_{34}}\right]\\
=\,&\frac{i}{2} \Big[ -2\times\pic{0.4}{bubble-s23-cut}+ (s_{12}+s_{13})\times\pic{0.4}{triangle-2m-s23-cut1} - s_{23} \times \pic{0.4}{triangle-s23-cut1}+s_{13}s_{23}\times\pic{0.4}{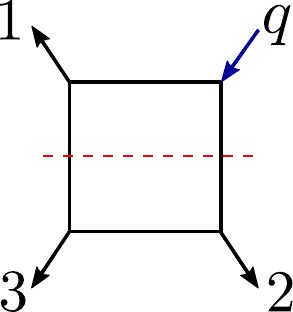}\Big]\ ,
\end{split}
\end{align}
which is identical to \eqref{eq:cut2a} apart from the box. Note that in the sum over
cyclic permutations of these two cuts three different one-mass boxes appear, each with their two possible
two-particle cuts. The cuts of the same boxes in the $q^2$-channel are already accounted for in \eqref{eq:q^2-cut-result}.
 
\subsubsection*{Diagram $\boldsymbol{(i)}$ $+$ Diagram $\boldsymbol{(ii)}$}
Combining the results \eqref{eq:cut2a} and \eqref{eq:cut2b} and noting that the coefficients of the integrals are consistent with those obtained from the $q^2$-channel cut in \eqref{eq:q^2-cut-result}, we find
\begin{align}
\label{eq:cut-q-p1-result}
\begin{split}
F_{\cO_F}^{(1)}(1^{\phi^{12}},2^{\phi^{23}},3^{\phi^{31}}&;q)\Big|_{2,s_{23}}\,=\,F_{\cO_F}^{(1)}(1^{\phi^{12}},2^{\phi^{23}},3^{\phi^{31}};q)\Big|_{2,s_{23},(i)}+F_{\cO_F}^{(1)}(1^{\phi^{12}},2^{\phi^{23}},3^{\phi^{31}};q)\Big|_{2,s_{23},(ii)}\\[5pt]
&=\, 
\frac{i}{2} \Big[-4\times\pic{0.4}{bubble-s23-cut}+ 2(s_{12}+s_{13})\times\pic{0.4}{triangle-2m-s23-cut1} - 2 s_{23}\times\pic{0.4}{triangle-s23-cut1}\\
&+s_{12}s_{23}\times\pic{0.4}{box-4-cut2}+s_{13}s_{23}\times\pic{0.4}{box-4-2-cut}\Big]\ .
\end{split}
\end{align}
Note that the coefficient of the box integral with $q$ inserted between $p_1$ and $p_3$ matches that obtained in the $q^2$-channel \eqref{eq:q^2-cut-result}, namely $(i/2)  (s_{12}s_{23})$. Moreover, the second box appearing in \eqref{eq:cut-q-p1-result} is detected in the $q^2$-cut with cyclically shifted external momenta: $1\rightarrow 2 \rightarrow 3 \rightarrow 1$.

\subsection{Final result}
Performing the cyclic sum we get the final result for the one-loop form factor:
\begin{align}
\begin{split}
&F^{(1)}_{\cO_F}(1^{\phi^{12}},2^{\phi^{23}},3^{\phi^{31}};q)\,=\, \frac{i}{2}\Big[ -4\times\pic{0.4}{bubble-s23}+ 2(s_{13}+s_{23})\times\pic{0.4}{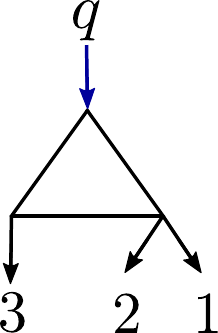} \\
& -2s_{23}\times\pic{0.4}{triangle-s23}+s_{12}s_{23}\times\pic{0.4}{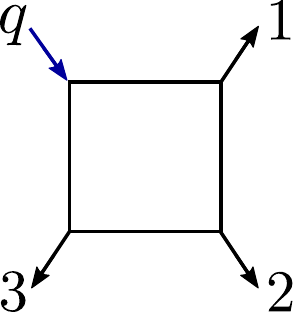}+\text{cyclic}(1,2,3)\Big]\ .
\end{split}
\end{align}
Expanding the result to $\O(\epsilon)$ we get
\begin{align}
\label{result:NonMin}
\begin{split}
F^{(1)}_{\cO_F}(1^{\phi^{12}},2^{\phi^{23}},3^{\phi^{31}};q)\,=\,&\frac{6}{\epsilon}+12+\frac{\pi^2}{2}-\Big[2\log(-s_{12})-\frac{1}{2}\log^2\frac{s_{12}}{s_{23}}\\
&-2\text{Li}_2\left(1-\frac{q^2}{s_{12}}\right)+\text{cyclic}(1,2,3)\Big]+\O(\epsilon)\ .
\end{split}
\end{align}
Importantly the infrared $1/\eps^2$ poles have cancelled in the final result, which is expected since the corresponding tree-level form factor does not exist. We can also rewrite the result using the variables $u$, $v$ ad $w$ introduced in \eqref{eq:uvw}, getting%
\footnote{Note that  under renormalisation this quantity will combine with   \eqref{rem-transc-2}.}
\beq
F^{(1)}_{\cO_F}(1^{\phi^{12}},2^{\phi^{23}}3^{\phi^{31}};q)\, =\,  
2 \, {\ \ (-s_{12})^{- \eps} \over \eps ( 1- 2 \eps) } \, 
-\, 
\Big[ 2\text{Li}_2 (1-u)  \, +\,  \log u \log v  \Big] \, + \,  \zeta_2  \, + \, \text{cyclic}(1,2,3) \ .
\eeq

\section{Two-loop sub-minimal form factor $\langle \bar\psi \bar\psi |\Tr X[Y,Z]| 0 \rangle$}
\label{sec-subminimal}

\begin{figure}[h]
\centering
\includegraphics[width=0.8\linewidth]{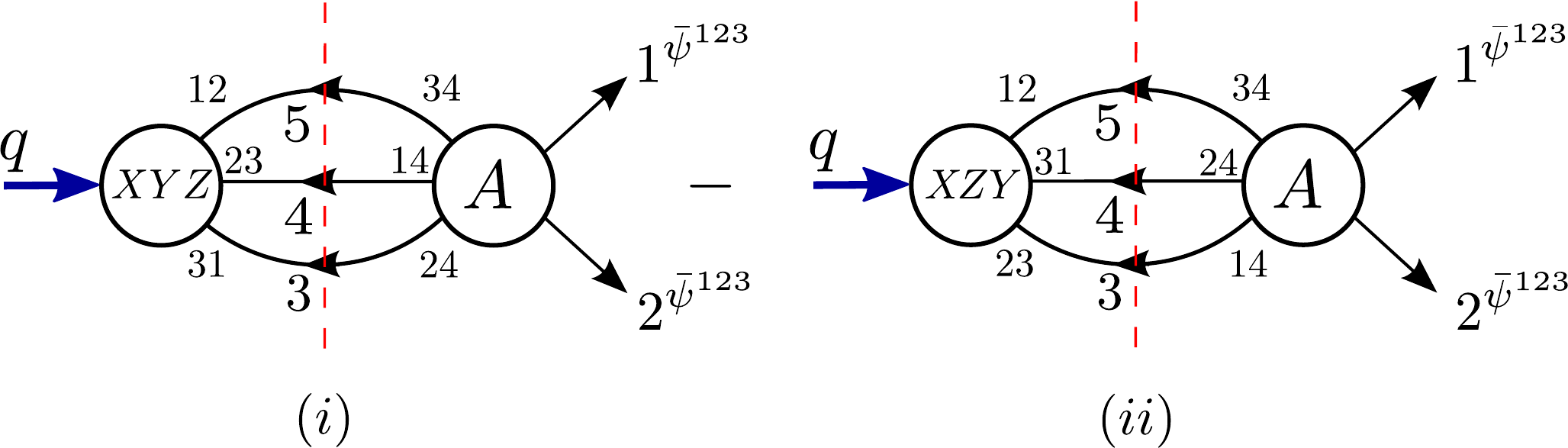}
\caption{\it  Triple cut of the two-loop sub-minimal form factor $\langle \bar\psi \bar\psi |\O_B| 0 \rangle$. The second set of identical diagrams, but with external legs 1 and 2 swapped has to be added, corresponding to the fact that it leads to the same colour-ordering.}
\label{fig:triple-cut-submin}
\end{figure}

Here we consider the second off-diagonal form factor in \eqref{eq:FF-matrix}, namely the sub-minimal form factor $\langle  \bar\psi \bar\psi  | \cO_B |0\rangle$ with $\cO_B= \Tr ( X[Y,Z])$ and $\langle \bar\psi \bar\psi|$ being a shorthand notation for $\langle 1^{\bar{\psi}^{123}} 2^{\bar{\psi}^{123}}|$.  As it is clear from Figure \ref{fig:triple-cut-submin},  this object exists only at two loops or more, hence we only need to consider the two three-particle cuts presented here.

For the first diagram, the relevant amplitude (and hence the integrand, since the tree-level form factor is just 1) is
\begin{align}
(i): A(1^{\bar{\psi}^{123}},2^{\bar{\psi}^{123}}, 3^{\phi^{24}}, 4^{\phi^{14}}, 5^{\phi^{34}}) &\,=\, -i\frac{\ls53\rs}{\ls 23\rs\ls 51\rs}
\,.
\end{align}
For the second diagram, the relevant amplitude is
\begin{align}
(ii): A(1^{\bar{\psi}^{123}},2^{\bar{\psi}^{123}}, 3^{\phi^{14}}, 4^{\phi^{24}}, 5^{\phi^{34}})\,=\, i\frac{\ls53\rs}{\ls 23\rs\ls 51\rs}\,,
\end{align}
which differs from $(i)$ only by a sign. Taking into account the relative minus sign between the two diagrams coming from the commutator and converting to momentum invariants we get 
\beqa\label{sub-cut-result}
(i)-(ii):\,\frac{1}{\ls 12 \rs} \cdot \frac{s_{35}s_{12}-s_{25}s_{13}+s_{15}s_{23}}{s_{23}s_{15}}\,,
\eeqa
where we have taken into account the factor of $i^3$ coming from the cut propagators.
We note that for the half-BPS case of $\tilde{\O}_{\rm BPS}=\mbox{Tr}(X\left\{Y,Z\right\})$ the two contributions would cancel out exactly, which is consistent with the fact that the operator is protected.

\noindent The cut integrand corresponding to the expression in (\ref{sub-cut-result}) is given by
\begin{align}
\label{eq:subminimal-before-reduction}
F^{(2)}_{\cO_B}(1^{\bar{\psi}^{123}},2^{\bar{\psi}^{123}};q)\Big|_{3,q^2}\,=\,\frac{1}{\ls12\rs}(s_{35}s_{12}-s_{25}s_{13}+s_{15}s_{23})\times \pic{0.5}{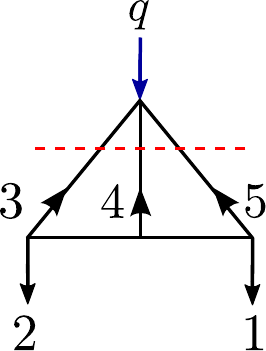}\ .
\end{align}
\noindent
Lifting the cut momenta off-shell and performing the integral reductions using the {\texttt{LiteRed}} package gives an $\epsilon$-dependent prefactor times a ``sunset" integral,
\begin{align}
F^{(2)}_{\O_B}(1^{\bar{\psi}^{123}},2^{\bar{\psi}^{123}};q)\,=\,\frac{1}{\ls12\rs}\frac{2(3\epsilon-2)}{2\epsilon-1}\times \pic{0.5}{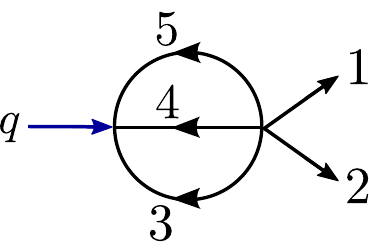}\ .
\end{align}
Note that any ambiguity associated with factors of $p_i^2,\,i=4,5,6$ in the numerator of \eqref{eq:subminimal-before-reduction} would lead to a (vanishing) scaleless integral.

Finally, we proceed to substitute the expression for the sunset  integral, which can be found in \cite{Gehrmann:1999as}. We also perform a summation over the cyclic permutations of the internal legs and note that having done so, the value of the five-point amplitude entering the cut does not change and so the result picks up an overall factor of three. Finally, a further factor of two is included corresponding to the two possible orderings of the external legs.

We proceed by expanding the results in powers of $\epsilon$  up to $\O(\epsilon)$ and get
\begin{align}
\label{subminresult}
\begin{split}
F^{(2)}_{\O_B}(1^{\bar{\psi}^{123}},2^{\bar{\psi}^{123}};q)\,=\,&
\frac{6}{\ls12\rs}\frac{\epsilon^2}{(1-2\epsilon)^2} \frac{\Gamma(1+2\epsilon) \Gamma(-\epsilon)^3 }{\Gamma(2 -3\epsilon)} (e^{ \gamma_{\rm E} \epsilon})^2 \left(-q^2\right)^{1-2\epsilon}
\\
\,=\,&
-6\b{12}\Big[\frac{1}{\epsilon}+ 7 -2 \log\left(-q^2\right)\Big] +\O(\epsilon)\, .
\end{split}
\end{align}
Note that this sub-minimal two-loop form factor has no lower-loop counterparts and,
therefore, it has only a $1/\epsilon$ UV divergence and no IR divergences.

\section{Two-loop dilatation operator in the $\boldsymbol{SU(2|3)}$ sector}
\label{mixing}

In this section we resolve the mixing  between the two  operators $\cO_B = \Tr  (X[Y, Z])$ and $\cO_F = (1/2) \Tr(\psi^\alpha \psi_\alpha)$ at two loops. Recall that all other dimension-three operators  in the $SU(2|3)$ sector such as  $\Tr (X^3)$, $\Tr (X^2 Y)$ and $\tilde{\O}_{\rm BPS} = \Tr (X \{ Y, Z\})$ are half-BPS and do not mix.  Doing so we will reproduce the two-loop dilatation operator for these operators in the $SU(2|3)$ sector originally derived in \cite{Beisert:2003ys}. 

\noindent We introduce the renormalised operators 
\begin{align}\label{MixMatrix}
\begin{pmatrix}
\O_F^{\rm ren} \\[10pt]
\O_B^{\rm ren} 
\end{pmatrix}\,=\,\begin{pmatrix}
\cZ_F^{\phantom{F}F} & \cZ_F^{\phantom{F}B} \\[10pt]
 \cZ_B^{\phantom{B}F} & \cZ_B^{\phantom{B}B}
\end{pmatrix} \begin{pmatrix}
\O_F \\[10pt]
\O_B
\end{pmatrix} \ , 
\end{align}
where $\O_F$ and $\O_B$ are the bare operators that we used to compute form factors in earlier sections. The matrix of renormalisation constants $\cZ$, also called mixing matrix, is  determined by requiring the UV-finiteness of the form factors of the renormalised operators $\O_F^{\rm ren}$ and $\O_B^{\rm ren}$ with the external states $\langle \bar X \bar Y \bar Z|$ and $\langle \bar\psi \bar\psi |$. 
The quantum correction to the dilatation operator $\mathfrak{D}$, denoted by $\delta \mathfrak{D}$, is related to the mixing matrix $\cZ$ as
\begin{align}
\label{eq:delta-D}
\delta\mathfrak{D}\,=\,\lim_{\eps\to 0}\Big[-\mu_R\frac{\partial}{\partial \mu_R} \log(\cZ)\Big] \,, 
\end{align}
where  $\mu_R$ is the renormalisation  scale. The relevant form factors from which we extract the renormalisation constants are written below, and we also indicate the schematic form of their perturbative expansions, as reflected by perturbative calculations: 
\beqa
\label{schematic1}
\langle \bar\psi \bar\psi  | \cO_F | \, 0\,  \rangle\Big|_{\rm UV}  & := &\ \langle 21 \rangle \Big[ f^{(1)} a(\mu_R)  +f^{(2)} a^2(\mu_R) + \cdots \Big]
\ ,
\\[5pt]
\label{schematic2}
 \langle \bar{X}\bar{Y}\bar{Z}  | \cO_F | \,  0 \,  \rangle\Big|_{\rm UV} &:=&  a(\mu_R)    \, \big[ g \cdot  h  \big] + \cdots \ , 
\\[5pt]
\label{schematic3}
\langle \bar\psi \bar\psi  | \cO_B | \, 0\,  \rangle\Big|_{\rm UV} &:= &\langle 21 \rangle a^2 (\mu_R)\,  \left( {1\over g} \cdot k\right)   + \cdots 
\ ,
\\[5pt]
\label{schematic4}
\langle \bar{X}\bar{Y}\bar{Z}  | \cO_B | \,  0 \,  \rangle\Big|_{\rm UV} & := &\ \  b^{(1)}  a (\mu_R)   \, + \, 
b^{(2)}  a^2 (\mu_R) + \cdots 
\ , 
\eeqa
where the coefficients carrying the UV divergences are
\beqa
f^{(1)}  &=&  {f^{(1)}_1 \over \epsilon}, \qquad \qquad f^{(2)} =  {f^{(2)}_2 \over \epsilon^2} + {f^{(2)}_1 \over \epsilon}\ , \nonumber 
\\
b^{(1)}  &=&  {b^{(1)}_1 \over \epsilon}, \qquad \qquad \ b^{(2)} =  {b^{(2)}_2 \over \epsilon^2} + {b^{(2)}_1 \over \epsilon}\ , \nonumber \\
h &=&{h_1\over \epsilon}  \, , \qquad \qquad \ \   \ k \ = \  {k_1\over \epsilon} \ , 
\eeqa
and the running 't Hooft coupling $a (\mu_R)$ defined in \eqref{Hooft-run} counts the number of loops. We have also been careful in distinguishing the coupling constant $g$ from $a(\mu_R)$ on the right-hand side of  \eqref{schematic1}--\eqref{schematic4}. Note that in \eqref{schematic1} and \eqref{schematic3} we have factored out the tree-level form factor $\la 1^{\bar\psi}2^{\bar\psi}| \frac{1}{2}{\rm Tr} (\psi \psi) | 0 \ra^{(0)}  = \la 21 \ra$.  

A few comments  on these expansions are in order. 

\begin{itemize}
\item[{\bf 1.}] We have performed explicit perturbative calculations in previous sections that allow us to extract \eqref{schematic2}--\eqref{schematic4}, and we will shortly explain how to extract the UV-poles for \eqref{schematic1}. 

\item[{\bf 2.}]
\eqref{schematic2} is the result of a one-loop calculation (hence the single power of $a(\mu_R)$ involving a five-point amplitude, which is $\cO (g^3)$ (hence the extra power of $g$).

\item[{\bf 3.}]  
\eqref{schematic3} is the result of a two-loop calculation, again involving a five-point amplitude. This is proportional to 
 $a(\mu_R)^2 / g$, which is $\cO (g^3)$ just like \eqref{schematic2}.
  
\end{itemize}

\noindent Expanding the mixing matrix $\cZ$ as
\begin{equation}
\label{eq:Z}
\cZ\,=\,\uno+\sum_{L=1}^{\infty}\cZ^{(L)} \ := \ \uno+\sum_{L=1}^{\infty} a({\mu_R})^L  z^{(L)}
 \ , 
\end{equation}
and requiring the finiteness of the renormalised form factors  we arrive at
\beqa
(z^{(1)})_F^{\phantom{B}F}   &=& -   {f^{(1)}_1 \over \epsilon}, \qquad \qquad (z^{(2)})_F^{\phantom{B}F}  =  -   {f^{(2)}_2 -(f^{(1)}_1)^2 \over \epsilon^2} -  {f^{(2)}_1 \over \epsilon}\ , \nonumber 
\\
(z^{(1)})_B^{\phantom{B}B}     &=&   - {b^{(1)}_1 \over \epsilon}, \qquad \qquad \ (z^{(2)})_B^{\phantom{B}B} =  - {b^{(2)}_2 -(b^{(1)}_1)^2 \over \epsilon^2}- {b^{(2)}_1 \over \epsilon} \ , \nonumber \\
(z^{(1)})_F^{\phantom{B}B}  &=& -  g \cdot {h_1\over \epsilon}  \, , \qquad \qquad \ \   \ (z^{(2)})_B^{\phantom{B}F}  \ = \   
- {1\over g} \cdot {k_1\over \epsilon} \ .
\eeqa
Note that from \eqref{eq:Z} $\cZ^{(L)} := a({\mu_R})^L z^{(L)}$.
The $ \log (\cZ)$ matrix has the form, up to $\cO\big( a(\mu_R)^2\big)$, 
{\footnotesize
\begin{align}
\begin{split}
\label{eq:log-Z}
&\log(\cZ)\sim
\begin{pmatrix}
(\mathcal{Z}^{(1)})_F^{\phantom{B}F}\!+\! \Big[(\mathcal{Z}^{(2)})_F^{\phantom{B}F}\!-\!\frac{1}{2}\big((\mathcal{Z}^{(1)})_F^{\phantom{B}F}\big)^2\Big] &  (\mathcal{Z}^{(1)})_F^{\phantom{B}B}-\frac{1}{2}(\mathcal{Z}^{(1)})_F^{\phantom{B}B}\Big[(\mathcal{Z}^{(1)})_F^{\phantom{F}F}+(\mathcal{Z}^{(1)})_B^{\phantom{B}B}\Big]\\[15pt]
 \, (\mathcal{Z}^{(2)})_B^{\phantom{B}F} & (\mathcal{Z}^{(1)})_B^{\phantom{B}B} \!+\! \Big[(\mathcal{Z}^{(2)})_B^{\phantom{B}B}\!-\!\frac{1}{2}\big((\mathcal{Z}^{(1)})_B^{\phantom{B}B}\big)^2\Big]
\end{pmatrix}
\\ \cr
=&
\begin{pmatrix}
- a(\mu_R)  \dfrac{f_1^{(1)}}{\epsilon}  - a^2 (\mu_R)\left[ \dfrac{f_2^{(2)} - {\frac{1}{2}} ( f_1^{(1)})^2 }{\epsilon^2}+\dfrac{f_1^{(2)}}{\epsilon}\right]  &  -g a (\mu_R)\cdot \dfrac{h_1}{\epsilon}\\[15pt]
 \, - \dfrac{a^2(\mu_R)}{g} \cdot \dfrac{k_1}{\epsilon} &-a(\mu_R)  \dfrac{b_1^{(1)}}{\epsilon}  - a^2 (\mu_R)\left[ \dfrac{b_2^{(2)} - {\frac{1}{2}} ( b_1^{(1)})^2 }{\epsilon^2}+\dfrac{b_1^{(2)}}{\epsilon}\right] 
\end{pmatrix}
\ .
\end{split}
\end{align}
}

We note that the term proportional to $(\mathcal{Z}^{(1)})_F^{\phantom{B}B}\Big[(\mathcal{Z}^{(1)})_F^{\phantom{F}F}+(\mathcal{Z}^{(1)})_B^{\phantom{B}B}\Big]$ is of order $ga^2(\mu_R)$, which is not relevant for operator mixing up to two loops and, hence, we drop it in going from the first to the second line of \eqref{eq:log-Z}. 

\noindent
We now move on to determine the various matrix elements.  From  \eqref{eq:XYZ-one-loop}  we  read off that 
\begin{align}
b_1^{(1)} = - 6\,,
\end{align}
and hence 
%$(z^{(1)})_B^{\phantom{B}B}$:
\begin{align}
(z^{(1)})_B^{\phantom{B}B}\,=\, \frac{6}{\eps}\ .
\end{align}
Next we compute $(z^{(2)})_B^{\phantom{B}B} - (1/2)( (z^{(1)})_B^{\phantom{B}B})^2$.  This quantity has already been calculated   in Section \ref{sect:TwoLoopRemainder}, and we  remark  that  we should drop the $\pi^2$ term  in \eqref{pisquared},   which is  not of UV origin. Doing so we find
\begin{align}
b^{(2)}_2=18\,,\qquad b_1^{(2)}=18\,,\qquad b^{(2)}_2 - (1/2) (b_1^{(1)})^2 = 0\,,
\end{align}
 and therefore 
\begin{align}
\label{ccccccc}
(z^{(2)})_B^{\phantom{B}B}-\frac{1}{2}((z^{(1)})_B^{\phantom{B}B})^2\,=\,-\frac{b_1^{(2)}}{\epsilon}\,=\, - \frac{18}{\eps} 
\ .
\end{align}
Importantly, the  $1/\eps^2$ pole  is absent in \eqref{ccccccc}. 
Next, from  the two-loop result of \eqref{subminresult}  we obtain $k_1=6$ and
\begin{align}
(z^{(2)})_B^{\phantom{B}F}\,=\,-\frac{6}{\eps}\cdot \frac{1}{g}
\ , 
\end{align}
while  from \eqref{result:NonMin} we find $h_1=6$ and
\begin{align}
(z^{(1)})_F^{\phantom{F}B} \,=\,-\frac{6}{\epsilon}\cdot g
\ .
\end{align}

Finally, we need to determine $(z^{(1)})_{F}^{\phantom{F}F}$ and $(z^{(2)})_{F}^{\phantom{F}F}$. In order to do so, we recall that 
$\cO_F$ appears as a component of the chiral part of the stress tensor multiplet operator (see Eq.~(3.3) of \cite{1103.3714}). Super form factors of this protected operator were first studied in \cite{Brandhuber:2011tv}. The components of this multiplet can be  obtained by acting with 
four of the eight supercharges $Q_{A\alpha}$ with $A=3,4$ on the bottom component $\Tr (X^2) = \Tr(\phi_{12}^2)$. Using the explicit supersymmetry transformation in Eqn.~(A.15) of \cite{1103.3714}, adapted to our conventions, and acting with $Q^\alpha_{3} Q_{3 \alpha}$ on the bottom component we find the following half-BPS descendent of $\Tr(\phi_{12}^2)$, 
\begin{align}
\label{aaaa}
\cO_{\rm BPS'}\,  := \, \frac{1}{2} \Tr (\psi^\alpha \psi_\alpha) \, + \, g \, \Tr (X [Y,Z])\,  =\,  \cO_F \, + \, g \, \cO_B\ .
\end{align}
Since this operator is half-BPS the corresponding form factors are UV finite. Hence we infer
that
\beq
\label{eq:relation-fermion-subminimal}
F_{\cO_F}(1^{\bar{\psi}^{123}},2^{\bar{\psi}^{123}};q)\Big|_{\rm UV}
=
-g F_{\cO_B}(1^{\bar{\psi}^{123}},2^{\bar{\psi}^{123}};q)\Big|_{\rm UV} ,
\eeq
from which we get 
\beq
(z^{(1)})_{F}^{\phantom{F}F} \,= - g (z^{(1)})_{B}^{\phantom{F}F} = \,0 \ ,\qquad (z^{(2)})_{F}^{\phantom{F}F} \,= - g (z^{(2)})_{B}^{\phantom{F}F} = \frac{6}{\epsilon} \ .
\eeq
Using \eqref{eq:relation-fermion-subminimal} we then obtain 
\begin{align}
 (z^{(2)})_F^{\phantom{B}F}-\frac{1}{2}\big((z^{(1)})_F^{\phantom{B}F}\big)^2\,=\, 
 %\textcolor{blue}{A} \ 
 \dfrac{6}{\epsilon} \ .
\end{align}
We can now write down  the matrix \eqref{eq:log-Z}, with the result
\begin{align}
%\label{eq:log-Z}
\log(\cZ)\,=\,\begin{pmatrix}
a^2 (\mu_R) \,\dfrac{6}{\epsilon}  &{-  a (\mu_R)\, g  \,\dfrac{6}{\eps}}\\[15pt]
 { - \dfrac{a^2 (\mu_R)}{g} \cdot \dfrac{6}{\eps}} &\quad  a (\mu_R)  \cdot  \dfrac{6}{\eps} \,  - \, a^2 (\mu_R) \cdot  \dfrac{18}{\eps}
\end{pmatrix}+ \cO\big( a(\mu_R)^3\big)\ .
\end{align}
Finally, the dilatation operator up to two loops is
\begin{align}
\label{twoldo}
&\delta \cD \ = \ \lim_{\eps\rightarrow 0}\Big[-\mu_R\frac{\partial}{\partial \mu_R}\log(\cZ)\Big]  \,=\,12\times 
\begin{pmatrix}
 2a^2 &   -a \, g  \\[15pt]
-2 \, \dfrac{a^2}{g} & a -  6\,  a^2
\end{pmatrix}\ , 
\end{align}
where we recall that our 't Hooft coupling is defined in \eqref{Hooft}. 
The eigenvalues of this matrix are the anomalous dimensions of the eigenstates of the dilatation operator. One of them vanishes indicating the presence of a non-trivial additional protected operator. The second one is 
\beq
\gamma_{\cK} \ = \ 12 \, a\,  -\,  48 \, a^2\,  + \cO(a^3)
\ , 
\eeq
in precise agreement with the one- and two-loop anomalous dimensions for the Konishi supermultiplet. 
We can also write the  corresponding eigenstates by diagonalising the transpose of $\delta \cD$.%
\footnote{Note that in this sector $\delta \cD$ is not symmetric. A generic combination of the two operators $\cO_F$ and $\cO_B$ can be written as $v_f \cO_F + v_b \cO_B := (  \mathbf{v},  \mathbf{O})$, with $\mathbf{v}^T := (v_F, v_B)$ and $\mathbf{O}^T := (\cO_F, \cO_B)$. Under the action of the dilatation operator we have  $(  \mathbf{v},  \mathbf{O}) \to 
( \mathbf{v},  \delta \cD \, \mathbf{O})  =  \big( (\delta \cD)^T \mathbf{v}, \mathbf{O}\big)$.} 
One arrives at the two operators \cite{Intriligator:1999ff, Bianchi:2001cm, Eden:2005ve, Eden:2009hz}
\beqa
\label{bpsother}
\cO_{\rm BPS'}^{\rm ren} & = &\ \cO_F^{\rm ren}  \, + \, g\, \cO_B^{\rm ren}\ , 
\\
\label{konid}
\cO_{\cK}^{\rm ren} & =& \cO_B^{\rm ren} \, - \, {g N \over 8 \pi^2} \, \cO_F^{\rm ren} \ . 
\eeqa
The first one is the protected operator introduced  in \eqref{aaaa} above, 
while the second combination is a descendant of the  Konishi operator.

\section{Conclusions}
\label{theend}

There are several natural continuations of the work presented in this paper.
In particular, it would be interesting to consider wider classes of non-protected operators than those  considered here and in \cite{Loebbert:2015ova,Nandan:2014oga}. Potentially this could lead to new insights and approaches to integrability. For example,  \cite{Wilhelm:2014qua} established a direct link between minimal one-loop form factors of general operators and Zwiebel's form of the one-loop dilatation operator \cite{Zwiebel:2011bx}. In \cite{Brandhuber:2015dta}
it was shown, using this form of the dilatation operator, how the Yangian symmetry \cite{Drummond:2009fd} of the tree-level $S$-matrix of $\cN\!=\!4$ SYM  implies the Yangian symmetry   of the one-loop dilatation operator, which in turn is related to its integrability \cite{Dolan:2003uh}. 
Clearly it would be very interesting to generalise this to higher loops.

In \cite{Brandhuber:2011tv}, supersymmetric  Ward identities were used to relate form factors of all the different operators in the protected stress tensor multiplet to form factors of the chiral primary operator $\Tr (X^2)$ at any loop order. This led naturally to the definition of super form factors extending the Nair on-shell superspace used for amplitudes in $\cN\!=\!4$ SYM.
It would be interesting to extend this to non-protected operators contained in larger multiplets. Technically this is more challenging but first important steps in this direction have been taken in recent papers \cite{Koster:2016ebi, Koster:2016loo} and \cite{1605.01386}
where tree-level MHV form factors for arbitrary unprotected operators were constructed using twistor actions and Lorentz harmonic chiral superspace, respectively.

It seems plausible that a more detailed study of minimal and slightly non-minimal two-loop form factors of non-protected operators will reveal a set of unique building blocks with different degrees of transcendentality for form factors of arbitrary operators. One piece of evidence
is the equivalence of the two-loop, three-point form factor of $\Tr (X^2)$ and the maximally transcendental part of Higgs to three-gluon amplitudes \cite{Brandhuber:2012vm}. It would be natural to expect that the universality of the leading transcendental part extends also to all length-two operators such as $\Tr ( DF DF)$ in any non-abelian gauge theory.
Another piece of evidence is that the minimal two-loop form factor of $\Tr (X^3)$ \cite{Brandhuber:2014ica} equals the
leading transcendentality part of the minimal two-loop form factors in the $SU(2)$ sector \cite{Loebbert:2015ova} and in the $SU(2|3)$ sector studied in the present paper. We would expect that this universality also applies to operators like $\Tr(F^3)$ in $\cN\!=\!4$ SYM, and possibly also in QCD and pure Yang-Mills.
Furthermore the intriguing relation of terms of lower transcendentality appearing in the 
$SU(2)$ and $SU(2|3)$ sectors (see Section \ref{sect:TwoLoopRemainder}) points at further unexpected regularities to be explored. We intend to return to these issues in the very near  future.

\section*{Acknowledgements}

We would like to thank  Burkhard Eden, Joseph Hayling, Paul Heslop, Gregory Korchemsky, Paolo Mattioli, Costis Papageorgakis, Sanjaye Ramgoolam, Radu Roiban and 
Rodolfo Russo  for useful discussions and comments. AB and GT would like to thank NORDITA and the organisers of the ``Aspects of Amplitudes" programme for hospitality during the last stage of this work. MK is supported by an STFC  studentship.  This work was supported by the Science and Technology Facilities Council Consolidated Grant ST/L000415/1  {\it ``String theory, gauge theory \& duality". } 

\newpage 

\appendix

\section{One-loop integral functions}\label{App:Integrals}

Throughout the paper, we use the following conventions for the one-loop massless scalar integrals in dimensional regularisation (upper/lower-case letters correspond to massive/massless momenta) \cite{Bern:1994cg}:
\begin{table}[h!]
\centering
\begin{tabular}{cl}
$\pic{0.6}{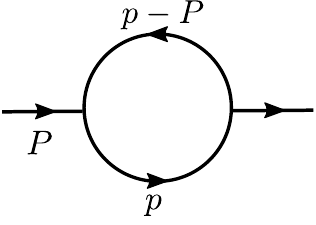}$& $ \,=\, \displaystyle\int  \dfrac{d^{4-2\epsilon}p}{(2\pi)^{4-2\epsilon}}\dfrac{1}{p^2(p-P)^2}  \,=\, \frac{i}{(4\pi)^{2-\epsilon}}\frac{r_{\Gamma}}{\epsilon(1-2\epsilon)}\left(-{P^2\over \mu^2}\right)^{-\epsilon}\ ,$\\[10pt]
$\pic{0.7}{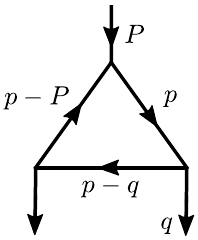}$ &$\,=\, \displaystyle \int \dfrac{d^{4-2\epsilon}p}{(2\pi)^{4-2\epsilon}}\dfrac{1}{p^2(p-q)^2(p-P)^2}= -\frac{i}{(4\pi)^{2-\epsilon}}\frac{r_\Gamma}{\epsilon^2}
%\dfrac{\left(-\dfrac{P^2}{\mu^2}\right)^{-\epsilon}}{- P^2}\ ,$ \\[10pt]
\dfrac{\ \ \left(-P^2 / \mu^2 \right)^{-\epsilon}}{(- P^2)}\ ,$ \\[10pt]
$\pic{0.7}{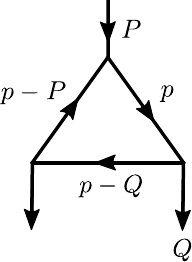}$ &$\,=\, \displaystyle \int \dfrac{d^{4-2\epsilon}p}{(2\pi)^{4-2\epsilon}}\dfrac{1}{p^2(p-Q)^2(p-P)^2}\,=\, -\frac{i}{(4\pi)^{2-\epsilon}}\frac{r_\Gamma}{\epsilon^2}\frac{\Big(-\dfrac{P^2}{\mu^2}\Big)^{-\epsilon}-\Big(-\dfrac{Q^2}{\mu^2}\Big)^{-\epsilon}}{(-P^2)-(-Q^2)}\ ,$
\\[10pt]
$\pic{0.8}{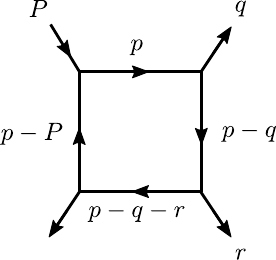} $ & $
\,=\, \displaystyle \int\! \dfrac{d^{4-2\eps} p}{(2\pi)^{4-2\eps}}\,\dfrac{1}{p^2 (p-q)^2(p-q-r)^2(p-P)^2}$ \\
& $ \displaystyle \,=\, -\frac{i}{(4\pi)^{2-\eps}}\frac{2r_{\Gamma}}{st}\Big\{-\frac{1}{\eps^2}\Big[\Big(-{s\over \mu^2}\Big)^{-\eps}+\Big(-{t\over \mu^2}\Big)^{-\eps} -\Big(-{P^2\over \mu^2}\Big)^{-\eps}\Big] $
\\[15pt]
& $\displaystyle \,+ \text{Li}_2\Big(1-\frac{P^2}{s}\Big) + \text{Li}_2\Big(1-\frac{P^2}{t}\Big) +\frac{1}{2}\log^2\Big(\frac{s}{t}\Big) + \frac{\pi^2}{6}\Big\}\ ,$
\end{tabular}
\end{table}

\noindent 
where
\begin{align*}
r_{\Gamma} \,=\, \frac{\Gamma(1+\epsilon)\Gamma(1-\epsilon)^2}{\Gamma(1-2\epsilon)}\,.
\end{align*}

\section{Comparing half-BPS form factors}\label{sect:compare-with-BPS}

In this appendix we present explicit calculations confirming that the minimal form factor of the half-BPS operator $\Tr(X\{Y,Z\})$  has  the same integrand, and hence remainder, as that of the minimal form factor  of $\Tr(X^3)$ considered in \cite{Brandhuber:2014ica}.

\begin{figure}[h]
\centering
\includegraphics[width=\linewidth]{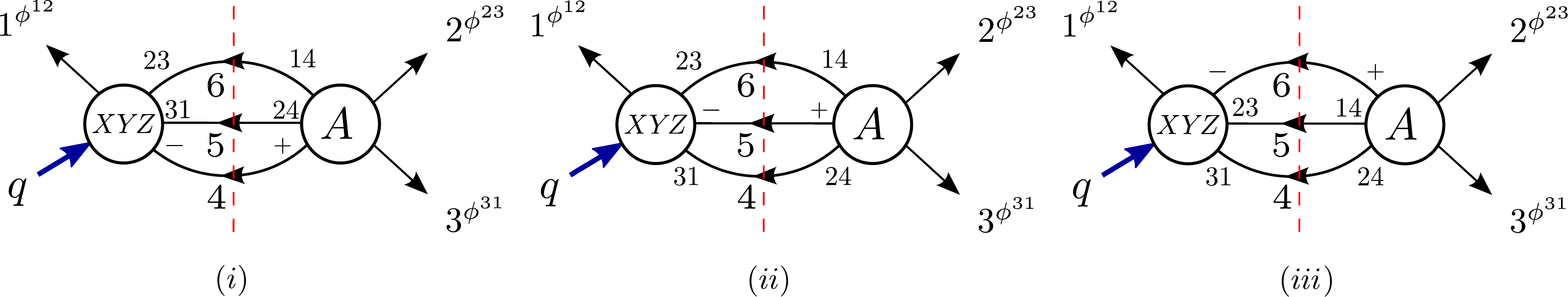}
\caption{\it Three possibilities for a single gluon running on one of the internal loop legs for $\Tr(XYZ)$ operator.}
\label{fig:XYZ-q+1-gluons-first}
\end{figure}

We begin by considering the three  diagrams in the gluonic contribution to the $s_{23}$-channel, presented in Figure \ref{fig:XYZ-q+1-gluons-first} below and corresponding to the $\Tr(XYZ)$ operator. 
The corresponding integrands are
\begin{align}
\begin{split}
(i):\ &i^3\, A(2^{\phi^{23}},3^{\phi^{31}}, 4^{+}, 5^{\phi^{24}}, 6^{\phi^{14}} )\times F_{\Tr(XYZ)}^{(0)}(1^{\phi^{12}}, 6^{\phi^{23}}, 5^{\phi^{31}}, 4^{-};q)\,=\,-\frac{\la 25\ra \la36\ra \ls 51\rs}{\la 34\ra \la45\ra\la 62\ra \ls 54\rs \ls41\rs}\,,\\
(ii):\ &i^3\, A(2^{\phi^{23}},3^{\phi^{31}}, 4^{\phi^{24}}, 5^{+}, 6^{\phi^{14}} )\times F_{\Tr(XYZ)}^{(0)}(1^{\phi^{12}}, 6^{\phi^{23}}, 5^{-}, 4^{\phi^{31}};q)\,=\, -\frac{\la 24\ra \la36\ra\la46\ra \ls 64\rs}{\la 34\ra \la45\ra\la56\ra\la 62\ra \ls 54\rs \ls65\rs}\,, \\
(iii):\ &i^3\, A(2^{\phi^{23}},3^{\phi^{31}}, 4^{\phi^{24}}, 5^{\phi^{14}}, 6^{+} )\times F_{\Tr(XYZ)}^{(0)}(1^{\phi^{12}}, 6^{-}, 5^{\phi^{23}}, 4^{\phi^{33}};q)\,=\, -\frac{\la 24\ra \la35\ra \ls 15\rs}{\la34\ra\la56\ra\la 62\ra \ls 16\rs \ls65\rs}\,.
\
\end{split}
\end{align}
We will combine these into ``anti-commutator" pieces by appropriately adding to them $-1/2$ of the terms that appear in expressions \eqref{E1}--\eqref{E3}, corresponding to the $\Tr(XZY)$ operator (removing the factor of $-2$). We then find   for the diagrams in Figure \ref{fig:XZY-q+1-gluons-first}$(i)$ and \ref{fig:XYZ-q+1-gluons-first}$(i)$, 
\begin{align}
\begin{split}
\rm{AC}_1 &\,=\, -\frac{\big(\la 25\ra \la36\ra+\la 35\ra \la62\ra\big) \ls 51\rs}{\la 34\ra \la45\ra\la 62\ra \ls 54\rs \ls41\rs} \,=\,-\frac{\la 23\ra \la56\ra \ls 51\rs}{\la 34\ra \la45\ra\la 62\ra \ls 54\rs \ls41\rs} \, .
\end{split}
\end{align}
%where we have used the Schouten identity
%\beqa
%\la ri\ra\la jk\ra+\la rj\ra\la ki\ra+\la rk\ra\la ij\ra \,=\,0\,.
%\eeqa
Similarly, we find for  the diagrams in   Figure \ref{fig:XZY-q+1-gluons-first}$(ii)$ and \ref{fig:XYZ-q+1-gluons-first}$(ii)$,
\beqa
\rm{AC}_2 &\,=\,& -\frac{\la 23\ra \la46\ra\la46\ra \ls 64\rs}{\la 34\ra \la45\ra\la56\ra\la 62\ra \ls 54\rs \ls65\rs}\,, 
\eeqa
and finally, for the integrands of Figure \ref{fig:XZY-q+1-gluons-first}$(iii)$ and \ref{fig:XYZ-q+1-gluons-first}$(iii)$,  
\beqa
\rm{AC}_3 &\,=\,& -\frac{\la 23\ra \la45\ra \ls 15\rs}{\la 34\ra\la56\ra\la 62\ra \ls 16\rs \ls65\rs}\,.
\eeqa
Next we consider the   fermionic contributions to this cut  for the operator $\Tr(XYZ)$. These are presented in Figure \ref{fig:XYZ-q+1-fermions} below.
\begin{figure}[h]
\centering
\includegraphics[width=0.8\linewidth]{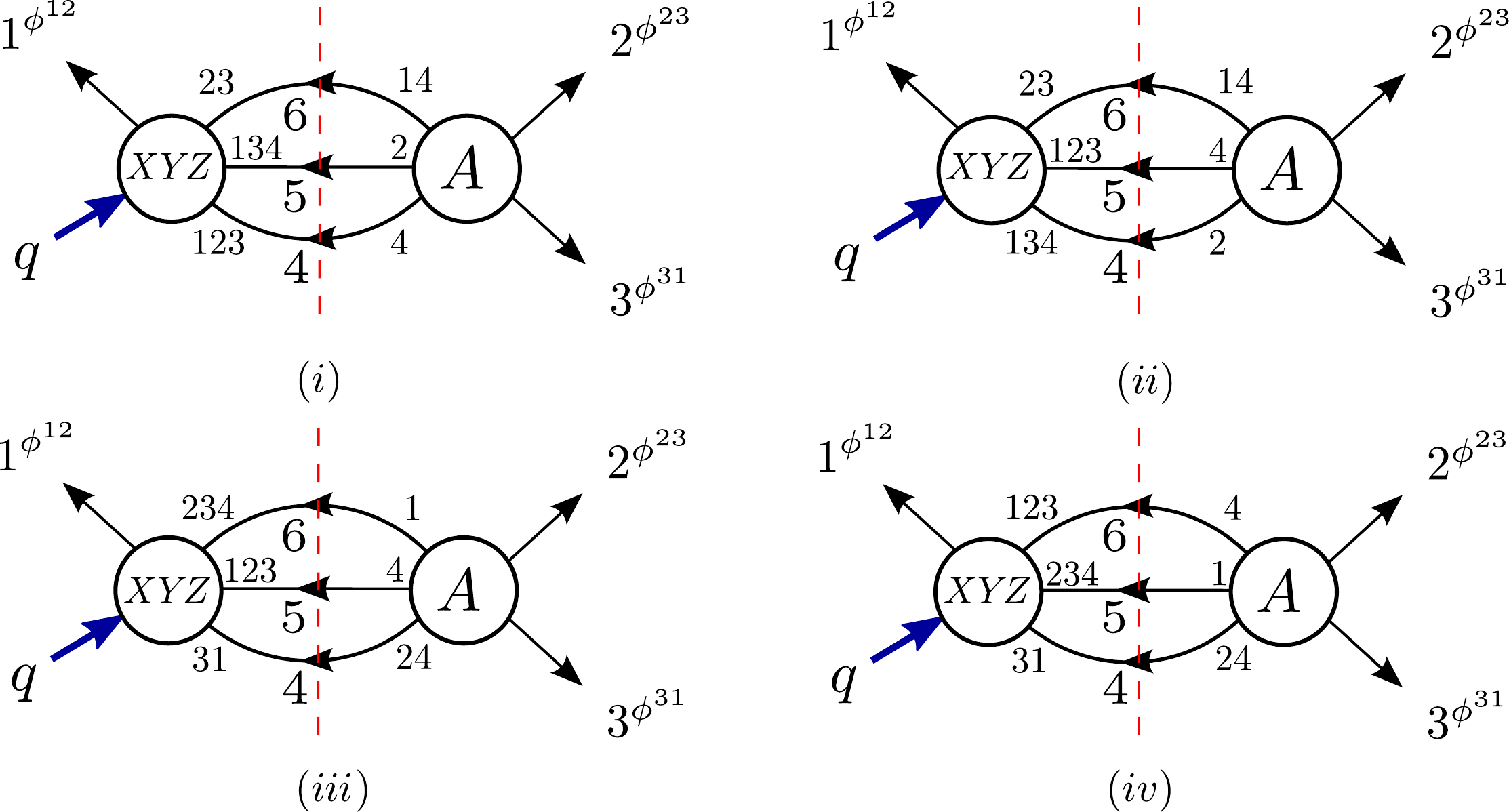}
\caption{\it Four possibilities for fermions running on the internal loop legs for the $\Tr(XYZ)$ operator.}
\label{fig:XYZ-q+1-fermions}
\end{figure}
The corresponding integrands are:
\begin{align}
\begin{split}
(i):\; i^3\,A(2^{\phi^{23}},3^{\phi^{31}}, 4^{\psi^{4}}, 5^{\psi^{2}}, 6^{\phi^{14}} )\times F_{\Tr(XYZ)}^{(0)}(1^{\phi^{12}}, 6^{\phi^{23}}, 5^{\psi^{134}}, 4^{\bar{\psi}^{123}};q)&\,=\,\frac{\la25\ra\la46\ra\la36\ra}{\la34\ra\la56\ra\la62\ra s_{45}}\ ,\\[5pt]
(ii):\; i^3\,A(2^{\phi^{23}},3^{\phi^{31}}, 4^{\psi^{4}}, 5^{\psi^{2}}, 6^{\phi^{14}} )\times F_{\Tr(XYZ)}^{(0)}(1^{\phi^{12}}, 6^{\phi^{23}}, 5^{\bar{\psi}^{123}}, 4^{\psi^{134}};q)&\,=\, -\frac{\la42\ra\la36\ra}{\la34\ra\la62\ra s_{45}}\ ,\\[5pt]
(iii):\; i^3\,A(2^{\phi^{23}},3^{\phi^{31}}, 4^{\phi^{24}}, 5^{\psi^{4}}, 6^{\psi^{1}} )\times F_{\Tr(XYZ)}^{(0)}(1^{\phi^{12}}, 6^{\psi^{234}}, 5^{\bar{\psi}^{123}}, 4^{\phi^{31}};q)&\,=\,-\frac{\la36\ra\la42\ra}{\la34\ra\la62\ra s_{56}}\,,\\[5pt]
(iv):\; i^3\,A(2^{\phi^{23}},3^{\phi^{31}}, 4^{\phi^{24}}, 5^{\psi^{1}}, 6^{\psi^{4}} )\times F_{\Tr(XYZ)}^{(0)}(1^{\phi^{12}}, 6^{\bar{\psi}^{123}}, 5^{\psi^{234}}, 4^{\phi^{31}};q)&\,=\, \frac{\la24\ra\la35\ra\la46\ra}{\la34\ra\la62\ra\la45\ra s_{56}}\, .\\[5pt]
\end{split}
\end{align}
We combine them similarly to the gluonic case: for the commutator, diagrams of Figure \ref{fig:q+1-fermions-first} and \ref{fig:q+1-fermions-second} should come with an overall minus sign. 
After some algebra we find,  for Figure \ref{fig:q+1-fermions-first} plus Figure \ref{fig:XYZ-q+1-fermions} $(i)$ and $(ii)$, 
\begin{align}
\rm{AC}_4 &\,=\, \frac{1}{\la34\ra\la56\ra\la62\ra s_{45}}\Big(\la25\ra\la36\ra\la46\ra -\la36\ra\la42\ra\la56\ra -\la26\ra\la35\ra\la46\ra +\la34\ra\la56\ra\la62\ra \Big)\nn
&\,=\, -\frac{\la23\ra}{\la34\ra\la45\ra\la56\ra\la62\ra}\frac{2\la46\ra\la65\ra}{\ls54\rs}\,,
\end{align}
and for Figure \ref{fig:q+1-fermions-second} plus Figure \ref{fig:XYZ-q+1-fermions} $(iii)$ and $(iv)$, 
\begin{align}
\rm{AC}_5 &\,=\, -\frac{1}{\la34\ra\la45\ra\la62\ra s_{56}} \Big(\la36\ra\la42\ra\la45\ra -\la24\ra\la35\ra\la46\ra +\la25\ra\la46\ra\la34\ra -\la34\ra\la45\ra\la62\ra \Big)\nn
&\,=\, -\frac{\la23\ra}{\la34\ra\la45\ra\la56\ra\la62\ra}\frac{2\la45\ra\la64\ra}{\ls65\rs}\,.
\end{align}
Finally  we combine  all  the ``anti-commutator" terms. After some manipulation, we get
\begin{align}\label{cutBPSsol}
\sum_{i=1}^5 {\rm AC}_i \,=\,-\frac{\la23\ra}{\la34\ra\la45\ra\la56\ra\la62\ra}\Big[&\frac{\ls51\rs\la54\ra^2}{\ls65\rs\ls16\rs} -2\frac{\la54\ra\la64\ra}{\ls65\rs} +\frac{\ls16\rs\la64\ra^2}{\ls65\rs\ls51\rs}
\nn &-\frac{\ls14\rs\la46\ra^2}{\ls45\rs\ls51\rs} +2\frac{\la46\ra\la56\ra}{\ls45\rs} -\frac{\ls51\rs\la56\ra^2}{\ls45\rs\ls14\rs}\Big]\,,
\end{align}
which is precisely the result of the $s_{23}$-channel cut of operator $\Tr(X^3)$ as presented in Eq.~(3.16) of \cite{Brandhuber:2014ica}. 

%\newpage

\bibliographystyle{utphys}
\bibliography{remainder}
\end{document}